\documentstyle[12pt,epsfig]{article}
\newskip\humongous \humongous=0pt plus 1000pt minus 1000pt

\newif\ifdtup

\def\ie{\hbox{\rm i.e.}{}} 
\def\eg{\hbox{\it e.g.}{}}

\def\abs#1{\left| #1\right|}
\def\pr#1{#1^\prime}

\def\beq{\begin{equation}}
\def\eeq{\end{equation}}

\def\beqn{\begin{eqnarray}}
\def\eeqn{\end{eqnarray}}
\relax

\def\dotx{\dotx{\dot\overline{x}}}

\relax

\catcode`\@=11

\@addtoreset{equation}{section}
\def\theequation{\thesection\arabic{equation}}

\def\@normalsize{\@setsize\normalsize{15pt}\xiipt\@xiipt
\abovedisplayskip 14pt plus3pt minus3pt%
\belowdisplayskip \abovedisplayskip
\abovedisplayshortskip \z@ plus3pt%
\belowdisplayshortskip 7pt plus3.5pt minus0pt}

\def\small{\@setsize\small{13.6pt}\xipt\@xipt
\abovedisplayskip 13pt plus3pt minus3pt%
\belowdisplayskip \abovedisplayskip
\abovedisplayshortskip \z@ plus3pt%
\belowdisplayshortskip 7pt plus3.5pt minus0pt
\def\@listi{\parsep 4.5pt plus 2pt minus 1pt
     \itemsep \parsep
     \topsep 9pt plus 3pt minus 3pt}}

\@twosidetrue

\relax

\catcode`@=12

\catcode`\@=11

\def\section{\@startsection{section}{1}{\z@}{3.5ex plus 1ex minus
   .2ex}{2.3ex plus .2ex}{\large\bf}}

\def\thesection{\arabic{section}.}

\def\appendix{\setcounter{section}{0}
 \def\thesection{APPENDIX \Alph{section}:}
 \def\theequation{\Alph{section}.\arabic{equation}}}

\def\ps@headings{\def\@oddfoot{}\def\@evenfoot{}
\def\@oddhead{\hbox{}\hfill
 \makebox[.5\textwidth]{\raggedright\ignorespaces --\thepage{}--
 \hfill {}}}  
\def\@evenhead{\@oddhead}
\def\subsectionmark##1{\markboth{##1}{}}
}

\ps@headings

\catcode`\@=12

\def\figcap{\section*{Figure Captions\markboth
 {FIGURECAPTIONS}{FIGURECAPTIONS}}\list
 {Fig. \arabic{enumi}:\hfill}{\settowidth\labelwidth{Fig. 999:}
 \leftmargin\labelwidth
 \advance\leftmargin\labelsep\usecounter{enumi}}}
 \relax
\def\tablecap{\section*{Table Captions\markboth
 {TABLECAPTIONS}{TABLECAPTIONS}}\list
 {Table \arabic{enumi}:\hfill}{\settowidth\labelwidth{Table 999:}
 \leftmargin\labelwidth
 \advance\leftmargin\labelsep\usecounter{enumi}}}
 \relax
\def\reflist{\section*{References\markboth
 {REFLIST}{REFLIST}}\list
 {[\arabic{enumi}]\hfill}{\settowidth\labelwidth{[999]}
 \leftmargin\labelwidth
 \advance\leftmargin\labelsep\usecounter{enumi}}}
 \relax

\catcode`\@=11

\def\ps@headings{\def\@oddfoot{}\def\@evenfoot{}
\def\@oddhead{\hbox{}\hfill
 \makebox[.5\textwidth]{\raggedright\ignorespaces --\thepage{}--
 \hfill {}}}    
\def\@evenhead{\@oddhead}
\def\subsectionmark##1{\markboth{##1}{}}
}

\ps@headings

\relax

\relax
\def\pl#1#2#3{{\it Phys. Lett. }{\bf #1}(19#2)#3}

\def\prl#1#2#3{{\it Phys. Rev. Lett. }{\bf #1}(19#2)#3}

\def\pr#1#2#3{{\it Phys. Rev. }{\bf #1}(19#2)#3}
\def\np#1#2#3{{\it Nucl. Phys. }{\bf #1}(19#2)#3}

\relax

\begin{document}            
\newcommand\sss{\scriptscriptstyle}
\newcommand\jetmsrex{dE_{\sss JT} d\eta_{\sss J} 
                     d\Omega_{\sss J}^{(2-2\ep)}}
\newcommand\jetmsr{d\mu_{\sss J}}
\newcommand\jetmsrf{dE_{\sss JT} d\eta_{\sss J} d\varphi_{\sss J}}
\newcommand\EJT{E_{\sss JT}}
\newcommand\mq{\mbox{$m_{\sss \rm Q}$}}
\newcommand\as{\alpha_{\sss S}}         
\newcommand\ep{\epsilon}
\newcommand\Stwo{{\cal S}_2}
\newcommand\Sthree{{\cal S}_3}
\newcommand\Stilde{\widetilde{{\cal S}}_3}
\newcommand\Qb{\overline{Q}}
\newcommand\qb{\overline{q}}
\newcommand\Js{{\sss J}}
\newcommand\Th{\theta}
\renewcommand\topfraction{1}       
\renewcommand\bottomfraction{1}    
\renewcommand\textfraction{0}      
\setcounter{topnumber}{5}          
\setcounter{bottomnumber}{5}       
\setcounter{totalnumber}{5}        
\setcounter{dbltopnumber}{2}       
\newsavebox\tmpfig
\newcommand\settmpfig[1]{\sbox{\tmpfig}{\mbox{\ref{#1}}}}
\def    \be             {\begin{equation}}
\def    \ee             {\end{equation}} 
\def    \ba             {\begin{eqnarray}}
\def    \ea             {\end{eqnarray}} 
\def    \nn             {\nonumber}
\def    \=              {\;=\;} 
\def    \frac           #1#2{{#1 \over #2}}
\def    \ret            {\\[\eqskip]}
\def    \eg             {{\em e.g.\/} }
\def \as   {\mbox{$\alpha_s$}}
\def \asopi{\mbox{$\frac{\alpha_s}{\pi}$}}
\def \oacube {\mbox{${\cal O}(\alpha_s^3)$}}
\def \oatwo {\mbox{${\cal O}(\alpha_s^2)$}}
\def \oas   {\mbox{${\cal O}(\alpha_s)$}}
\def \ppbar {\mbox{$p \bar p$}}         
\def \bbbar {\mbox{$b \bar b$}}         
\def \ccbar {\mbox{$c \bar c$}}         
\def \pt   {\mbox{$p_{\scriptscriptstyle T}$}}                
\def \et   {\mbox{$E_{\scriptscriptstyle T}$}}                
\def \etsq {\mbox{$E_{\scriptscriptstyle T}^2$}}                
\def \kt   {\mbox{$k_{\scriptscriptstyle T}$}}                
\def \rap   {\mbox{$\eta$}}
\def \deltar {\mbox{$R$}}
\def \mur  {\mbox{$\mu_{\rm \scriptscriptstyle{R}}$}}                
\def \muf  {\mbox{$\mu_{\rm \scriptscriptstyle{F}}$}}               
\def \muo  {\mbox{$\mu_0$}}                      
\def \to   {\mbox{$\rightarrow$}}
\def    \mb             {\mbox{$m_b$}}  
\def    \mc             {\mbox{$m_c$}}  

\begin{titlepage}
\nopagebreak
{\flushright{
        \begin{minipage}{4cm}
        CERN-TH/96-85  \hfill \\
        ETH-TH/96-09 \hfill \\
        hep-ph/9605270\hfill \\
        \end{minipage}        }

}
\vfill
\begin{center}
{\LARGE { \bf \sc  Heavy-Quark Jets \\[0.3cm]
           in Hadronic Collisions}}             
\vskip .5cm
{\bf Stefano FRIXIONE}\footnote{Work supported by the National Swiss 
Foundation.}
\\                    
\vskip .1cm
{Theoretical Physics, ETH, Zurich, Switzerland} \\
\vskip .5cm                                      
{\bf Michelangelo L. MANGANO\footnote{On leave of absence
from INFN, Pisa, Italy.}}
\\
\vskip 0.1cm
{CERN, TH Division, Geneva, Switzerland} \\
\end{center}
\nopagebreak
\vfill
\begin{abstract}
  We present a next-to-leading order QCD calculation of the production rates of
jets containing heavy quarks. This calculation is performed using the
standard Snowmass jet algorithm; it therefore allows a comparison with similar
results known at  next-to-leading order for generic jets. As an application, we
present results for the inclusive transverse energy of
charm and bottom jets  at the Tevatron collider, with a
complete study of the dependence on the jet cone-size and of the theoretical
uncertainties.                              
\end{abstract}        
\vskip 1cm
CERN-TH/96-85 \hfill \\
May 1996 \hfill
\vfill
\end{titlepage}
\section{Introduction}
The study of heavy quark production has provided some of
the most interesting results in the physics of high energy hadronic collisions.
The large available data sets of $b$-hadrons have been used for precise 
measurements of spectroscopy and lifetimes, as well as for measurements of the
production rates. The associated production of jets including $b$-hadrons and
$W$ vector bosons has been used for the detection of the $t$-quark. 
Several signals for new physics, such as an intermediate mass Higgs or the
supersymmetric partners of the top quark, could manifest themselves via the
presence in the final state, among other things, of jets containing $b$-quarks.
A close study of the production properties of $b$-jets in QCD is therefore an
important phenomenological input for many of these searches. Calculations have
been performed in the past for the production of $b$-quarks at next-to-leading
order (NLO) in QCD~[\ref{NDE},\ref{MNR}]. They have been used in comparisons
with data measured at the $\rm Sp\bar pS$~[\ref{ua1}] and 
Tevatron~[\ref{bdata}]
colliders. In this paper we present a calculation of the production rates of
jets including heavy quarks, at NLO in QCD. The main difference between the
study of a heavy quark and a heavy-quark jet is that in the former case one is
interested in the momentum of the quark itself, regardless of the properties of
the event in which the quark in embedded, while in the latter case one is
interested in the properties of a jet containing one or more heavy quarks,
regardless of the momentum fraction of the jet carried by the quark. A priori
it is expected that variables such as the \et\ distribution 
of a heavy-quark jet
should be described by a finite-order QCD calculation more precisely than the
\pt\ distribution of open quarks. This is because at high momentum large
logarithms $\log(\pt/m)$ appear at any order in the perturbative expansion of
the open quark \pt\ distribution, due to the emission of hard collinear gluons.
These logarithms need to be resummed, using techniques such as fragmentation
functions~[\ref{greco}]. Collinear logarithms, on the other hand, are not
present in the \et\ distribution of heavy-quark jets, 
since the jet \et\ does not depend on
whether the energy is carried all by the quark or is shared among the quark and
collinear gluons. The experimental measurement of the \et\ distribution of 
heavy-quark jets does not depend either 
on the knowledge of the heavy-quark fragmentation functions,
contrary to the case 
of the \pt\ distribution of open heavy quarks. 
Experimental systematics, such as
the knowledge of decay branching ratios for heavy hadrons or            
of their decay spectra are also largely reduced.
                                                                             
The contents of this paper are as follows: in section~2 we introduce our
notation, definitions, and summarize the technique used in the calculation.
Section~3 presents our phenomenological results, with a complete discussion of
the production rates and properties of charm and bottom jets at the Tevatron
collider. Our conclusions can be found in section~4, while a thorough
discussion of the technical details of the calculation is collected in
the Appendix.                                                               
\section{Cross sections}

In this section, we briefly discuss the definition
of the heavy-quark jet cross section in perturbative QCD.
The interested reader will find a more detailed
presentation in the Appendix.

Thanks to the factorization theorem~[\ref{CSS}], a cross section 
in hadronic collisions can be written in the following way
\beq
d\sigma^{(H_1 H_2)}(P_1,P_2)=\sum_{ab}\int dx_1 dx_2 f^{(H_1)}_a(x_1)
f^{(H_2)}_b(x_2)d\hat{\sigma}_{ab}(x_1 P_1,x_2 P_2)\,,
\label{factth}
\eeq
where $H_1$ and $H_2$ are the incoming hadrons, $P_1$ and $P_2$
their momentum, and the sum runs over all the parton flavours
which give a non-trivial contribution. The quantities
$d\hat{\sigma}_{ab}$ are the subtracted partonic short-distance 
cross sections, which are calculable in perturbative QCD.
Our aim is to evaluate these quantities for heavy-quark jet production.
We will deal with $d\sigma_{ab}$, which are directly related
to Feynman diagrams; $d\hat{\sigma}_{ab}$ are obtained from
$d\sigma_{ab}$ by subtracting some suitable counterterms
for the initial state collinear singularities~[\ref{CSS}].

In perturbative QCD, a jet has to be defined in terms of
unobservable partons. The definition can be rather freely chosen,
the only constraint being that it must guarantee the infra-red
safeness of the cross section. In the present paper we adopt
the Snowmass convention~[\ref{snowmass}], whereby particles are clustered in
cones of radius $R$ in the pseudorapidity-azimuthal angle plane.
                       
The calculation of the heavy-quark jet cross section is very similar to
the one of the generic-jet cross section, but two
important differences have to be stressed.
By its very definition, a heavy-quark jet is not flavour-blind; we 
have to look for those jets containing a heavy flavour. Furthermore, the
mass of the heavy flavour is acting as a cutoff against final
state collinear divergences; therefore, we will not have to deal
with many singular contributions, which are present in the 
generic-jet partonic cross section.

At the leading order, the heavy-quark jet cross section in hadronic
collisions gets contributions from the following partonic
processes
\beqn
g\,+\,g\,\,\rightarrow\,\,Q\,+\,\Qb,
\label{prc1}
\\
q\,+\,\qb\,\,\rightarrow\,\,Q\,+\,\Qb.
\label{prc2}
\eeqn
At the next-to-leading order, the cross section gets contributions
from the radiative correction to the processes~(\ref{prc1}) and ({\ref{prc2}),
and from the tree amplitudes of the processes                   
\beqn
g\,+\,g\,\,\rightarrow\,\,Q\,+\,\Qb\,+\,g,
\label{prc3}
\\
q\,+\,\qb\,\,\rightarrow\,\,Q\,+\,\Qb\,+\,g,
\label{prc4}
\\
q\,+\,g\,\,\rightarrow\,\,Q\,+\,\Qb\,+\,q.
\label{prc5}
\eeqn
In the following, we will consider the case in which the heavy-quark jet
contains the heavy quark. This is by no means restrictive; the
heavy antiquark obviously can be treated in the same way.
                          
We write the leading-order contribution to the heavy-quark jet cross
section as
\beq
d\sigma^{(0)}_{ab}=\Stwo\,{\cal M}_{ab}^{(0)}\,d\Phi_2\,d\mu_{\sss J}\,,
\label{LOxsec}
\eeq
where ${\cal M}_{ab}^{(0)}$ is the leading-order transition amplitude 
for the two-to-two process \mbox{$a+b \; \to \; Q+\Qb$}, squared and summed
over spin and colour degrees of freedom, and divided by the flux and
average factors; $d\Phi_2$ is the two-body phase space for the
$Q\Qb$ pair,
\beq
d\mu_{\sss J}\,=\,\jetmsrex
\label{jetmeasure}
\eeq
is the measure over jet variables in $4-2\ep$ dimensions, and $\Stwo$ is 
the so-called measurement function, which defines the jet variables
in terms of the partonic variables (although eq.~(\ref{LOxsec}) is
completely general, the explicit form of the 
measurement function depends upon the merging algorithm
adopted: for a discussion on the use of the measurement 
function in jet physics, see refs.[\ref{KS},\ref{FKS}]).
In eq.~(\ref{jetmeasure}) we indicated with $E_{\sss JT}$ the
transverse energy of the jet, and with $\eta_{\sss J}$ its 
pseudorapidity; $d\Omega_{\sss J}^{(2-2\ep)}$ is the angular measure
in $2-2\ep$ dimensions. The measurement function is
actually a distribution, since the jet variables are implicitly
defined as entries of some $\delta$ distributions; this is the reason
for the $d\mu_{\sss J}$ on the RHS of eq.~(\ref{LOxsec}).
The explicit form of $\Stwo$ is given in the Appendix.

The next-to-leading order contribution is
\beq
d\sigma^{(1)}_{ab}=d\sigma^{(v)}_{ab}+d\sigma^{(r)}_{ab}\,,
\label{NLOxsec}
\eeq
where
\beq
d\sigma^{(v)}_{ab}=\Stwo\,{\cal M}_{ab}^{(v)}\,d\Phi_2\,d\mu_{\sss J}
\label{virtxsec}
\eeq
is the virtual part, due to the radiative corrections to the
two-to-two processes, and
\beq
d\sigma^{(r)}_{ab}=\Sthree\,{\cal M}_{ab}^{(r)}\,d\Phi_3\,d\mu_{\sss J}
\label{realxsec}
\eeq
is the real part, due to the two-to-three processes. 
Here $d\Phi_3$ is the three-body phase space 
for the $Q\Qb$ pair plus a light parton, and $\Sthree$
is analogous to $\Stwo$, but takes into account the fact that, in
the final state, one additional parton is present (and therefore
the jet definition has to be suitably modified).

It is apparent that, at the leading order, the heavy-quark jet 
can only coincide with the heavy quark itself. Therefore, at
the leading order, the heavy-quark jet cross section is identical
to the open-heavy-quark cross section. At the next-to-leading
order, the presence of a light parton in the final state enriches
the kinematical structure. The heavy-quark jet can be the heavy quark,
or it can contain the heavy quark and the light parton, or 
the heavy quark and the heavy antiquark. The heavy-quark jet
cross section is therefore different from the open-heavy-quark one.
Nevertheless, thanks to the non-zero mass of the quark, the structure
of the singularities of the heavy-quark jet cross section is identical
to the one of the open-heavy-quark cross section; a detailed proof
of this statement is reported in the Appendix. We can then write
the heavy-quark jet cross section at the next-to-leading order,
\beq                                                           
d\sigma_{ab}=d\sigma^{(0)}_{ab}+d\sigma^{(1)}_{ab} \; ,
\label{fullxsec}                                       
\eeq
in the following way:
\beq                 
d\sigma_{ab}=d\sigma^{(open)}_{ab}+d\Delta_{ab}\,,
\label{xsecsplit}
\eeq
where $d\sigma^{(open)}_{ab}$ is the open-heavy-quark cross
section, and $d\Delta_{ab}$ is implicitly defined in 
eq.~(\ref{xsecsplit}). The Fortran code for $d\sigma^{(open)}_{ab}$
is available from the authors of ref.~[\ref{MNR}]. For
the present paper, we wrote a Fortran code for the
evaluation of $d\Delta_{ab}$.  

\newcommand{\ccaption}[2]{
  \begin{center}
    \parbox{0.85\textwidth}{
      \caption[#1]{\small\it {#2}}}
  \end{center}    }
                                   
\section{The structure of heavy-quark jets at the Tevatron}
As an application of the formalism developed so far, 
we present in this section some results of interest for measurements
at the Fermilab 1.8~TeV Tevatron $p\bar p$ Collider~[\ref{koehn}]. 
We will consider
jets containing either charm or bottom quarks. For these we 
will provide absolute predictions for the production rates as a
function of the jet transverse energy and jet cone size $R$, 
and will explore the theoretical                           
uncertainties associated to the choice of factorization (\muf) and
renormalization (\mur) scales. Since we will be considering jets of energy much
larger than the heavy-quark mass,  the uncertainty associated to the mass
values chosen for  charm and bottom quarks is negligible, and  will not be
discussed. We will also study the fraction of heavy-quark jets relative to
generic jets, and the fraction of 
$b$-jets relative to $c$-jets. For this particular distribution, we will show
that most of the uncertainties related to the choice of scales cancel out in
the ratio, leaving a rather accurate NLO theoretical prediction. 

We consider jets produced within $|\eta|<1$,
in order to simulate a realistic geometrical acceptance of the Tevatron
detectors. We will use the parton distribution set MRSA$^\prime$~[\ref{MRSAP}].
Our default values of the parameters entering the calculations are:     
\mc~=~1.5~GeV, \mb~=~4.75~GeV, $\mur^2=\muf^2=\muo^2\equiv m_Q^2+\etsq$ and
\deltar~=~0.7.         
          
\begin{figure}
\centerline{\epsfig{figure=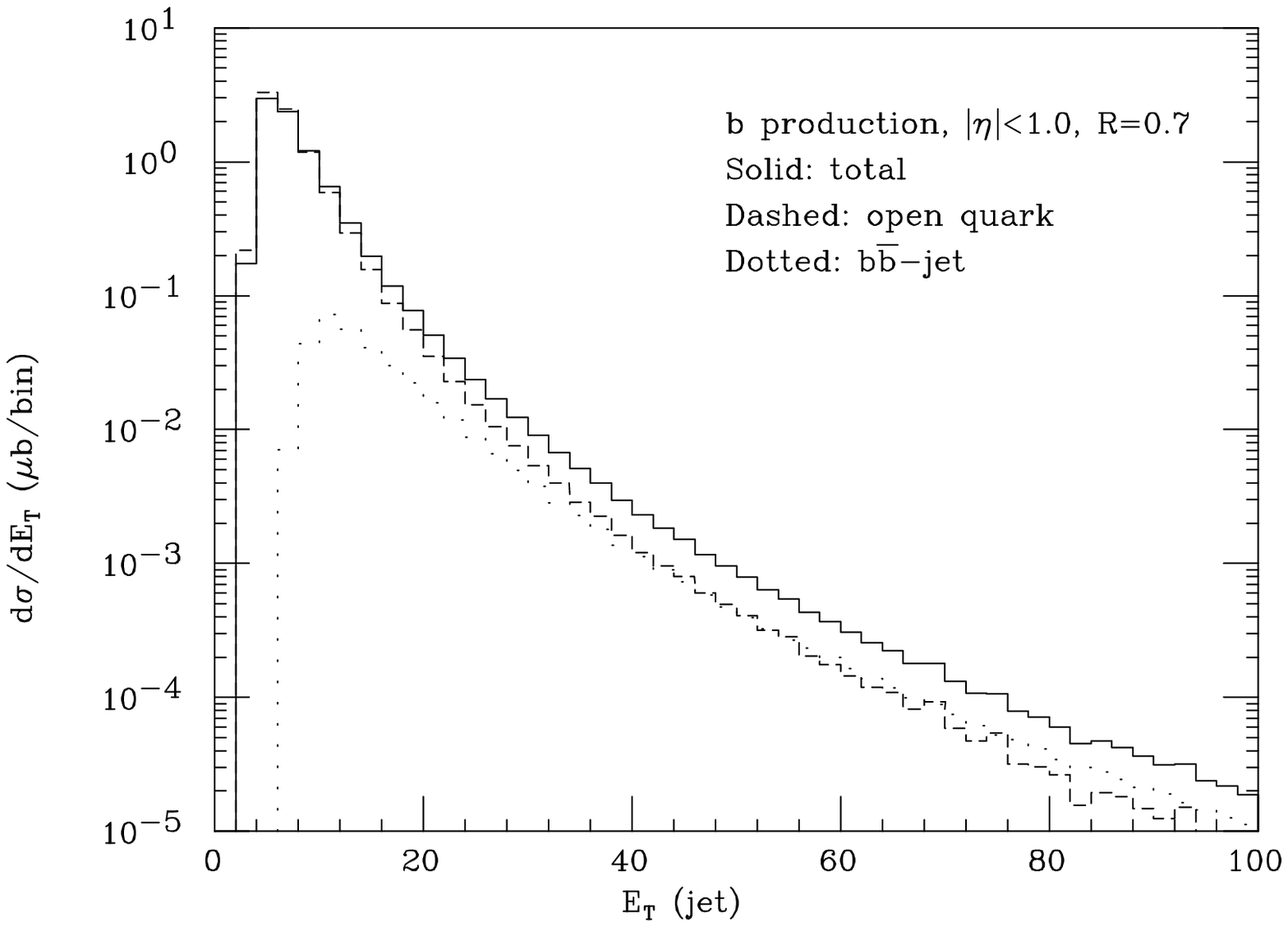,width=0.7\textwidth,clip=}}
\ccaption{}{ \label{fbinc}         
$b$-jet inclusive \et\ distribution in $p\bar p$ collisions at 1.8 TeV, for 
$|\eta|<1$, $R=0.7$ and $\muf=\mur=\muo$ (solid line). 
For comparison, we also show the open-quark inclusive
\et\ distribution (dashed line). The component of the jet-like
contribution due to jets containing both $b$ and $\bar b$ is represented by the
dotted line.}
\end{figure}

We start from the absolute production rates. Figure~\ref{fbinc} shows the
prediction for the \et\ distribution of $b$-jets at the Tevatron. For the
purpose of illustration, and to provide a direct estimate of the  
effects calculated in this paper, we separate in the figure the contribution of
the open-quark component. As indicated in the figure,
the jet-like component, defined as the additional contribution to
the open-quark one (the $\Delta$ term in eq.~(\ref{xsecsplit})), 
becomes dominant as soon as \et\ becomes
larger than 50~GeV. We also show the part of the jet-like component due to
jets that include the \bbbar\ pair (we will call these \bbbar\ jets). 
The figure suggests that for this \et\
range and with $R=0.7$ this  is the dominant part of the jet-like component.
This is consistent with the expectation that, for large enough \et\  and
provided  that the majority of the final-state generic  jets are composed of
primary gluons, heavy-quark jets are dominated by the process of gluon
splitting, with the jet formed by the heavy-quark pair. 
As we will show later, the situation changes at higher \et\ values, where
heavy quarks are mostly produced via the $s$-channel annihilation of light
quarks.                                  
                                                           
\begin{figure}
\centerline{\epsfig{figure=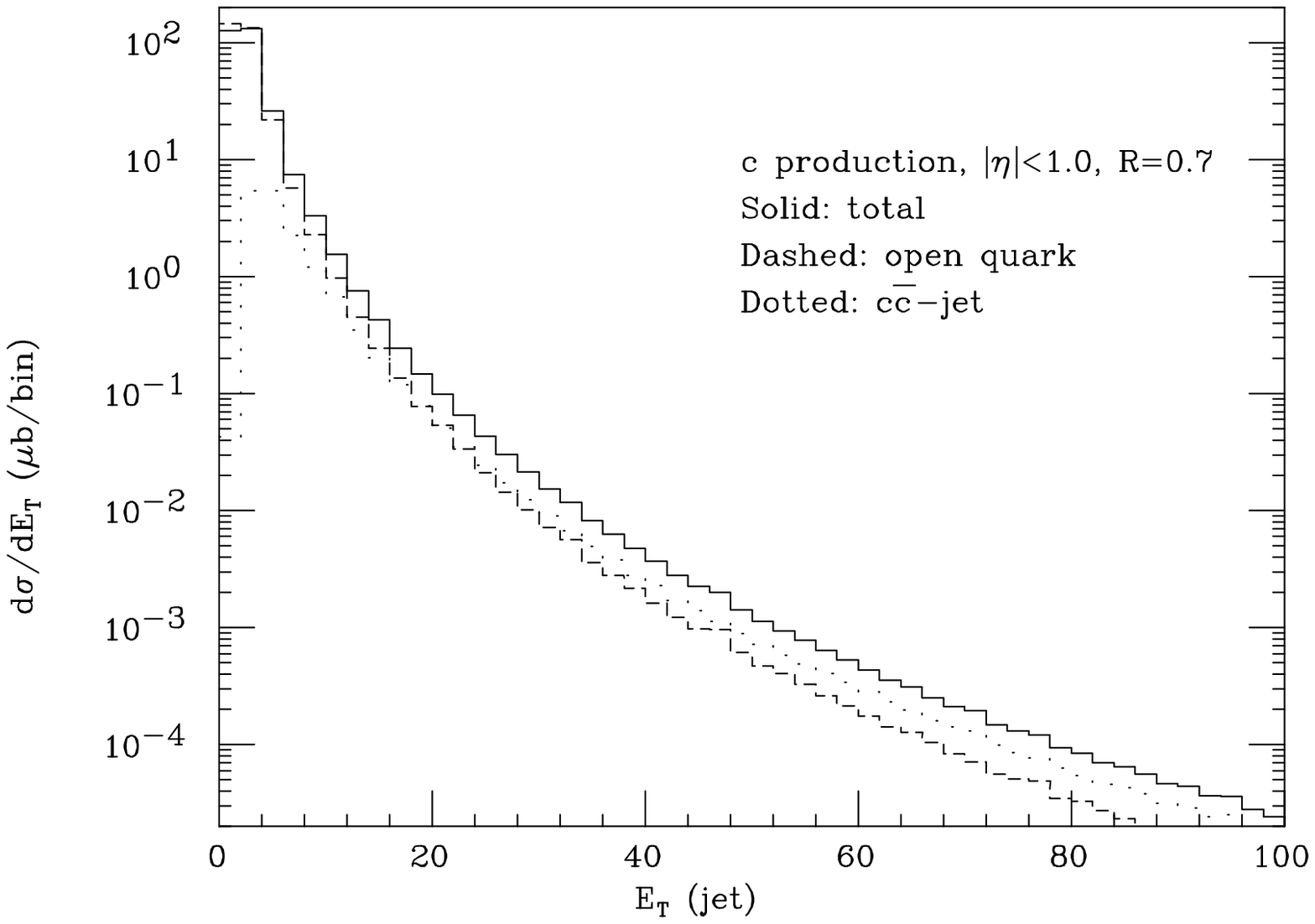,width=0.7\textwidth,clip=}}
\ccaption{}{ \label{fcinc}         
$c$-jet inclusive \et\ distribution in $p\bar p$ collisions at 1.8 TeV, for 
$|\eta|<1$, $R=0.7$ and $\muf=\mur=\muo$ (solid line). 
For comparison, we also show the open-quark inclusive
\et\ distribution (dashed line). The component of the jet-like
contribution due to jets containing both $c$ and $\bar c$ is represented by the
dotted line.}                                           
\end{figure}

Figure~\ref{fcinc} shows the same distributions for $c$-jets. Notice that the
value of \et\ at which the jet-like component becomes dominant is smaller than
in the case of $b$-jets. Again this is in agreement with naive expectations. 
The relative probability 
of finding a heavy-quark pair inside a high-\et\ gluon scales in fact like
$\log (\et/m_Q)$~[\ref{hvqjet}].                                 

\begin{figure}
\centerline{\epsfig{figure=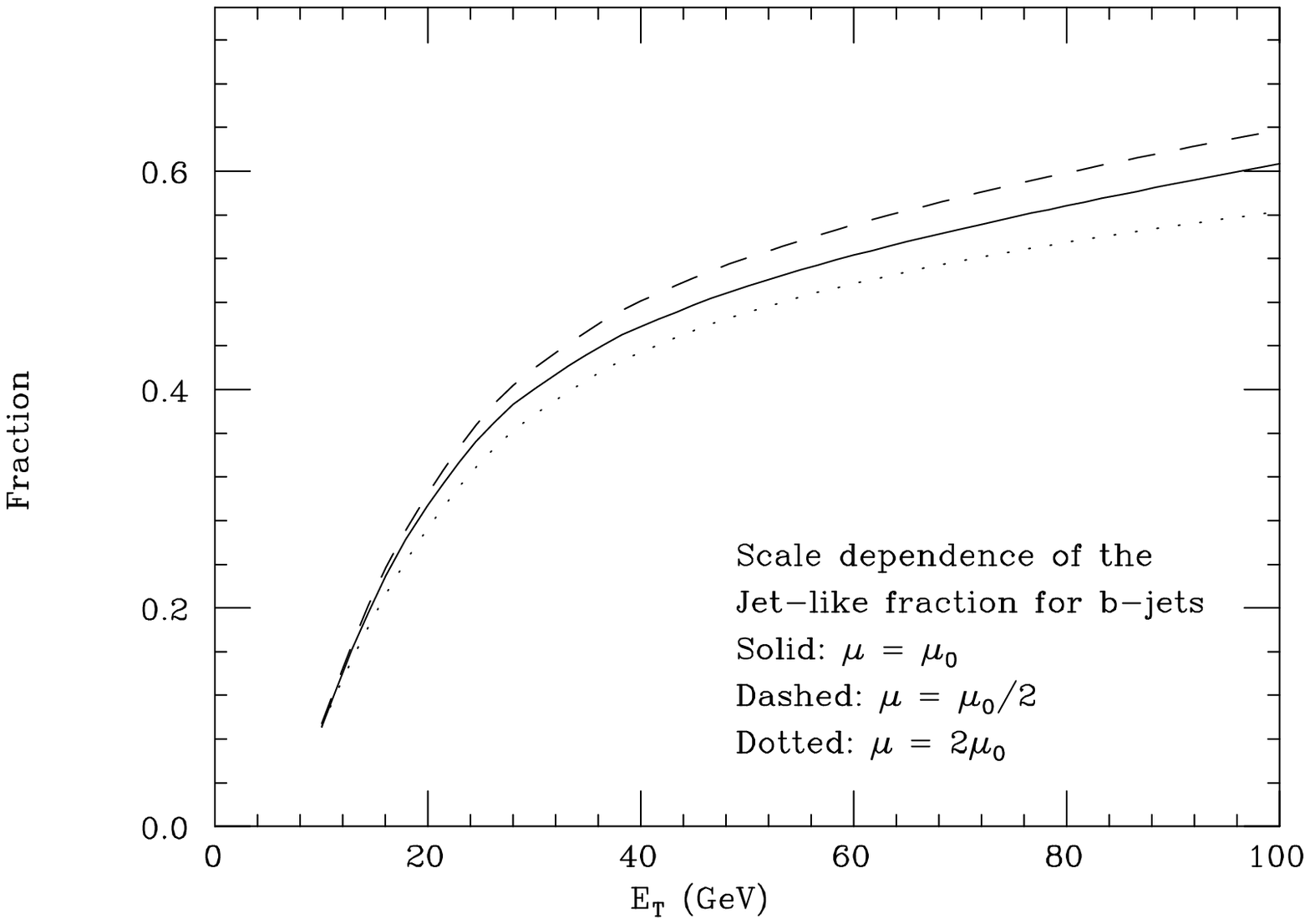,width=0.5\textwidth,clip=}
            \hspace{0.3cm}
            \epsfig{figure=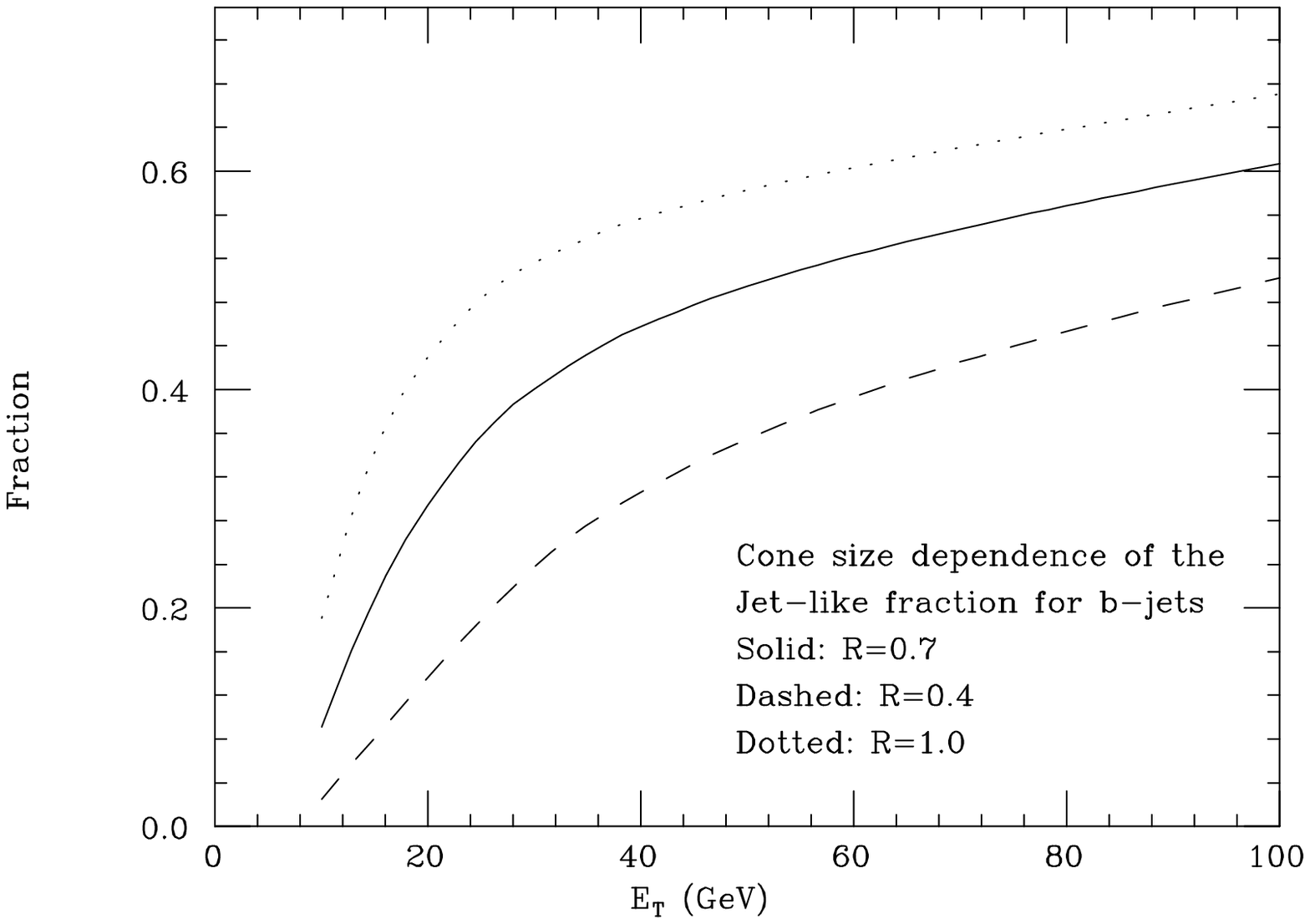,width=0.5\textwidth,clip=}}
\ccaption{}{ \label{fbfrac}                      
Left: relative contribution of the jet-like component in the 
$b$-jet inclusive \et\ distribution, for different values of the
renormalization and factorization scales ($\mur=\muf=\mu$).                    
Right: relative contribution of the jet-like component in the 
$b$-jet inclusive \et\ distribution, for various cone sizes $R$.}
\end{figure}                                                              
\begin{figure}
\centerline{\epsfig{figure=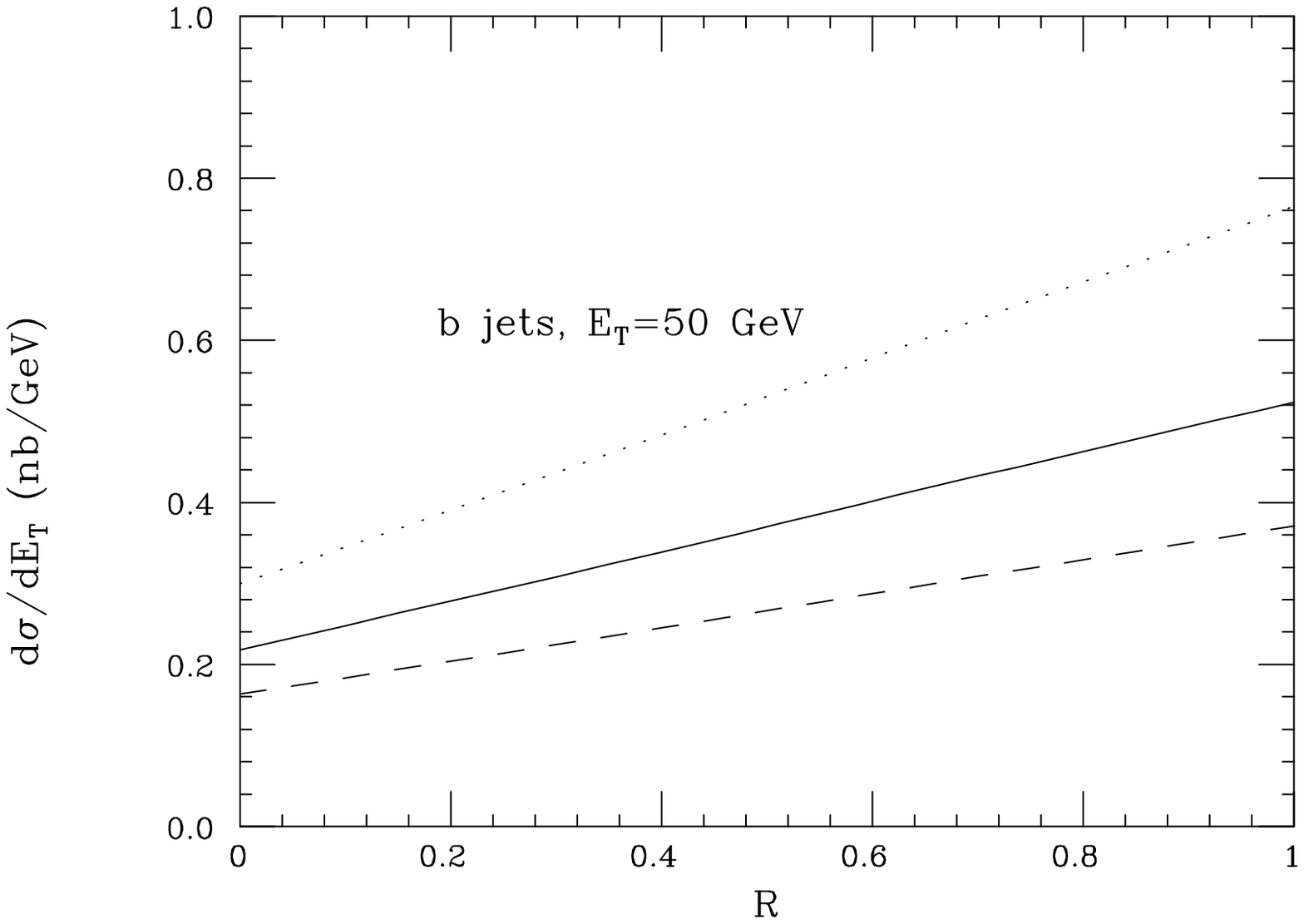,width=0.5\textwidth,clip=}
            \epsfig{figure=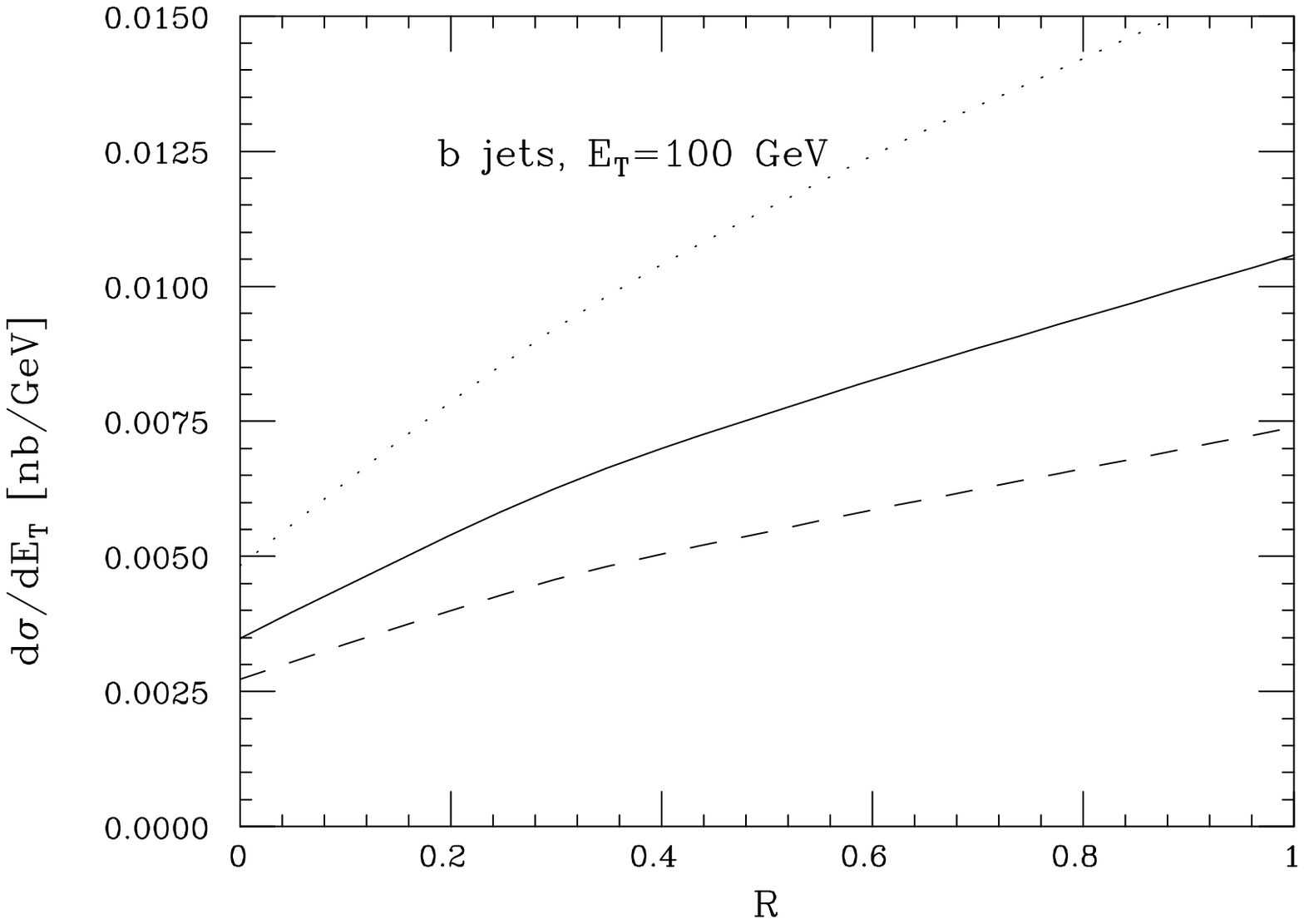,width=0.5\textwidth,clip=}}
\ccaption{}{ \label{fbcone}                      
$b$-jet inclusive \et\ rate, as a function of the cone size $R$, for 
\et~=~50~GeV (left) and \et~=~100~GeV (right).}
\end{figure}                                                              
Figure~\ref{fbfrac} shows the fraction of the jet-like component
  of $b$-jets versus \et, for various choices of 
  factorization/renormalization scale and cone size.
In the former case the dependence
on scale variations is rather small, while the dependence upon the jet
definition is more significant. Notice in particular that the \et\ value at
which the jet-like component becomes dominant depends significantly on the cone
size, being equal to 25, 50 and 100~GeV for $R=1$, 0.7 and 0.4 respectively.
\begin{figure}
\centerline{\epsfig{figure=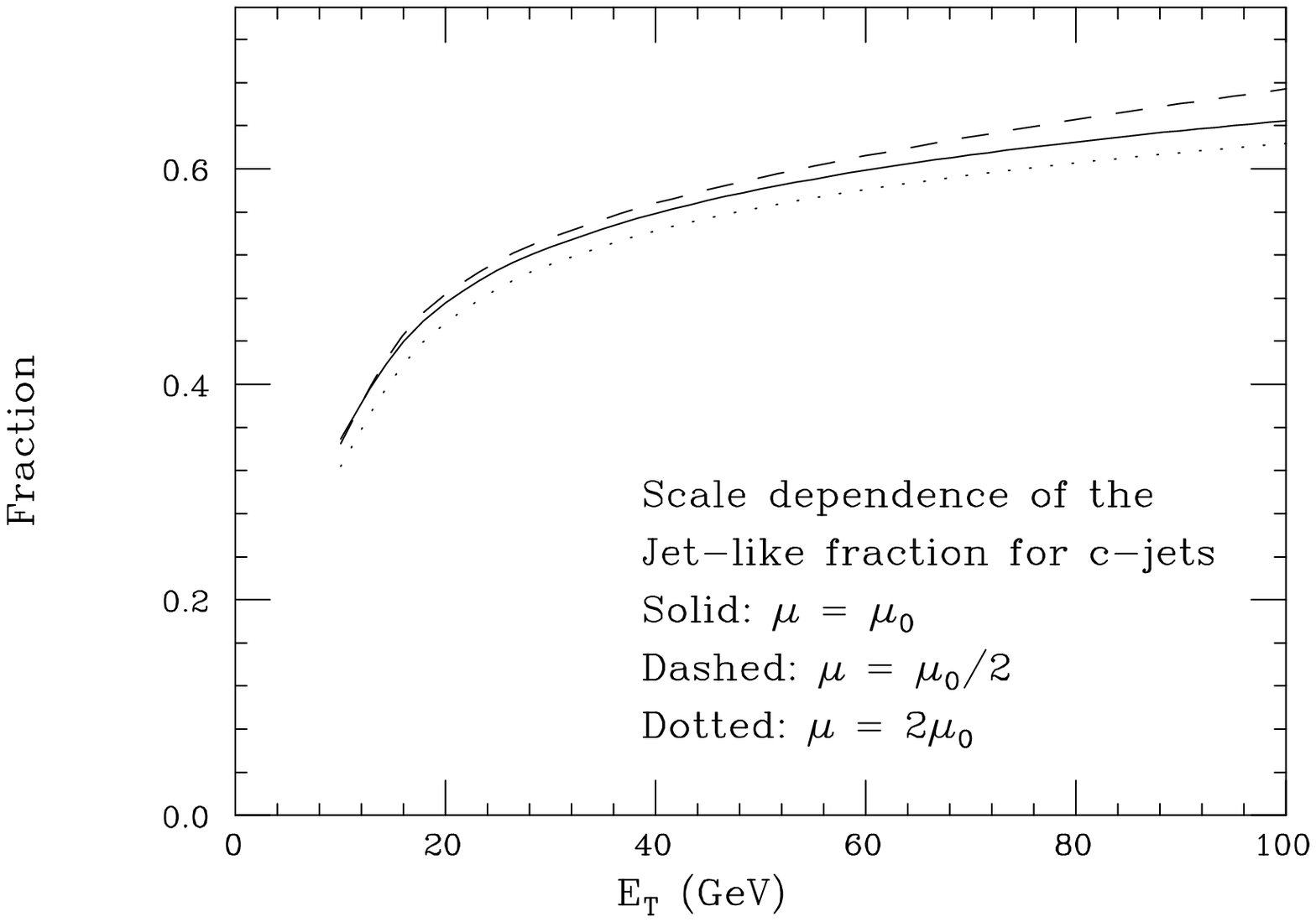,width=0.5\textwidth,clip=}
            \hspace{0.3cm}
            \epsfig{figure=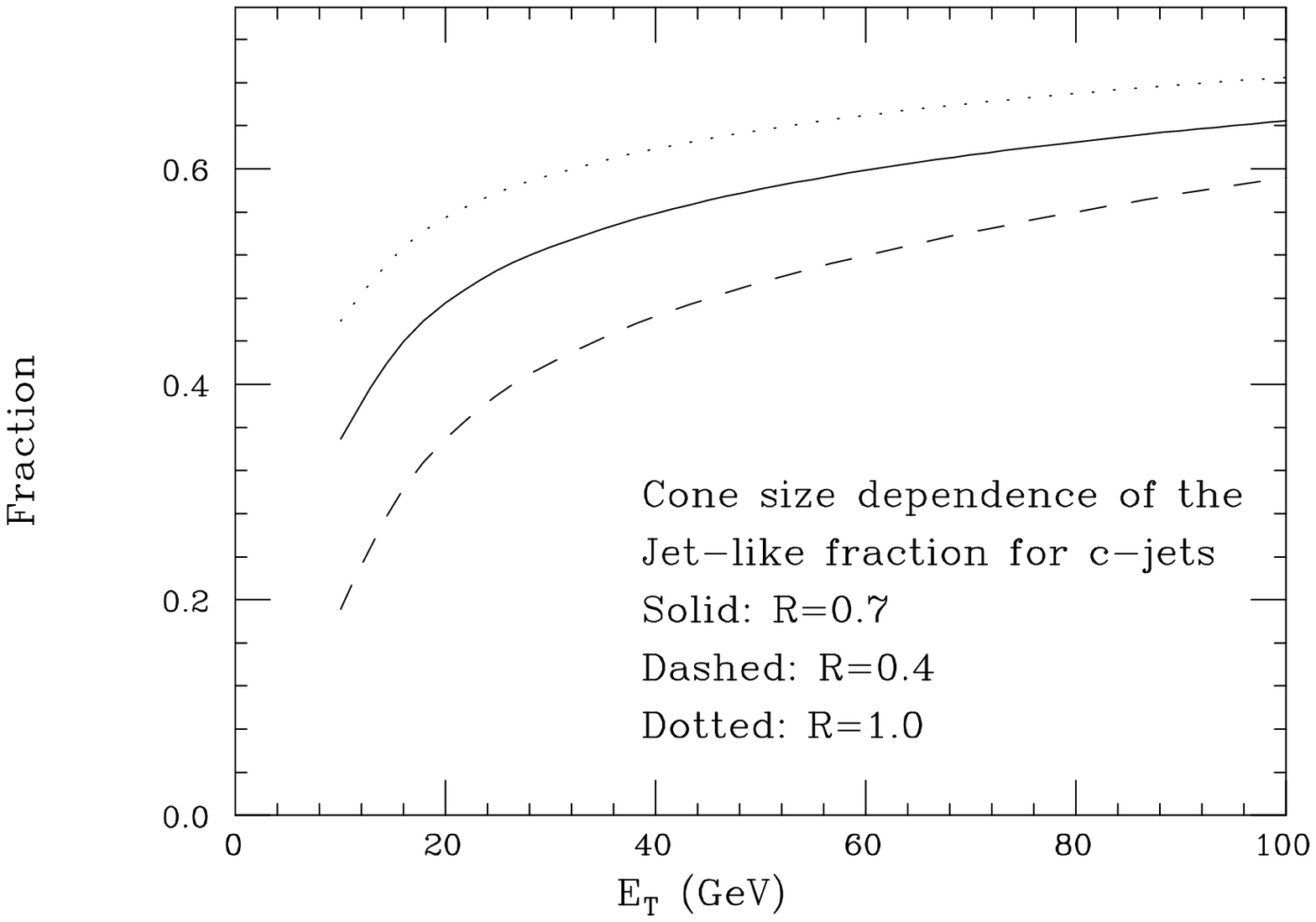,width=0.5\textwidth,clip=}}
\ccaption{}{ \label{fcfrac}                      
Left: relative contribution of the jet-like component in the 
$c$-jet inclusive \et\ distribution, for different values of the
renormalization and factorization scales ($\mur=\muf=\mu$).                    
Right: relative contribution of the jet-like component in the 
$c$-jet inclusive \et\ distribution, for various cones sizes $R$.}
\end{figure}                                                              
\begin{figure}
\centerline{\epsfig{figure=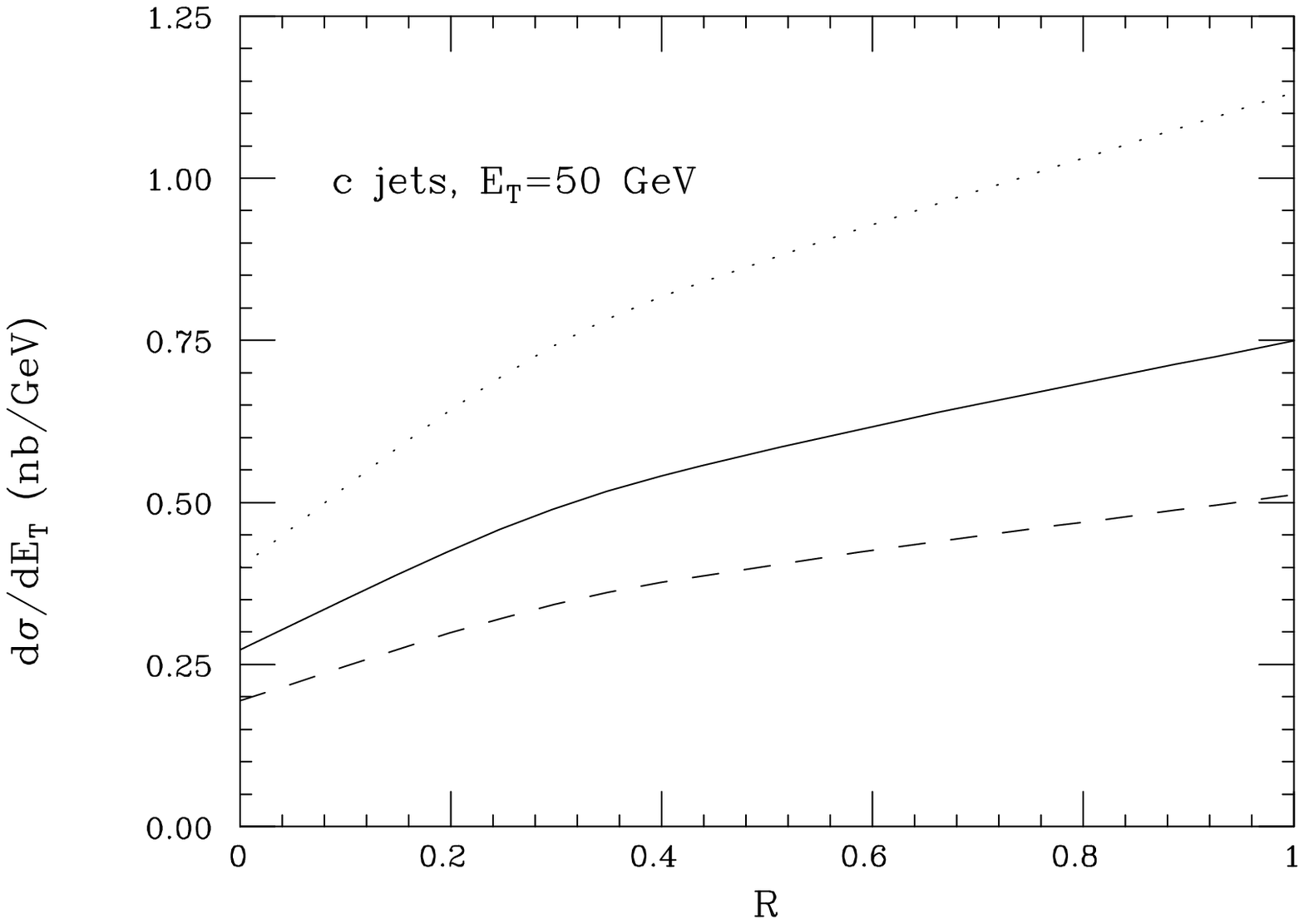,width=0.5\textwidth,clip=}
            \epsfig{figure=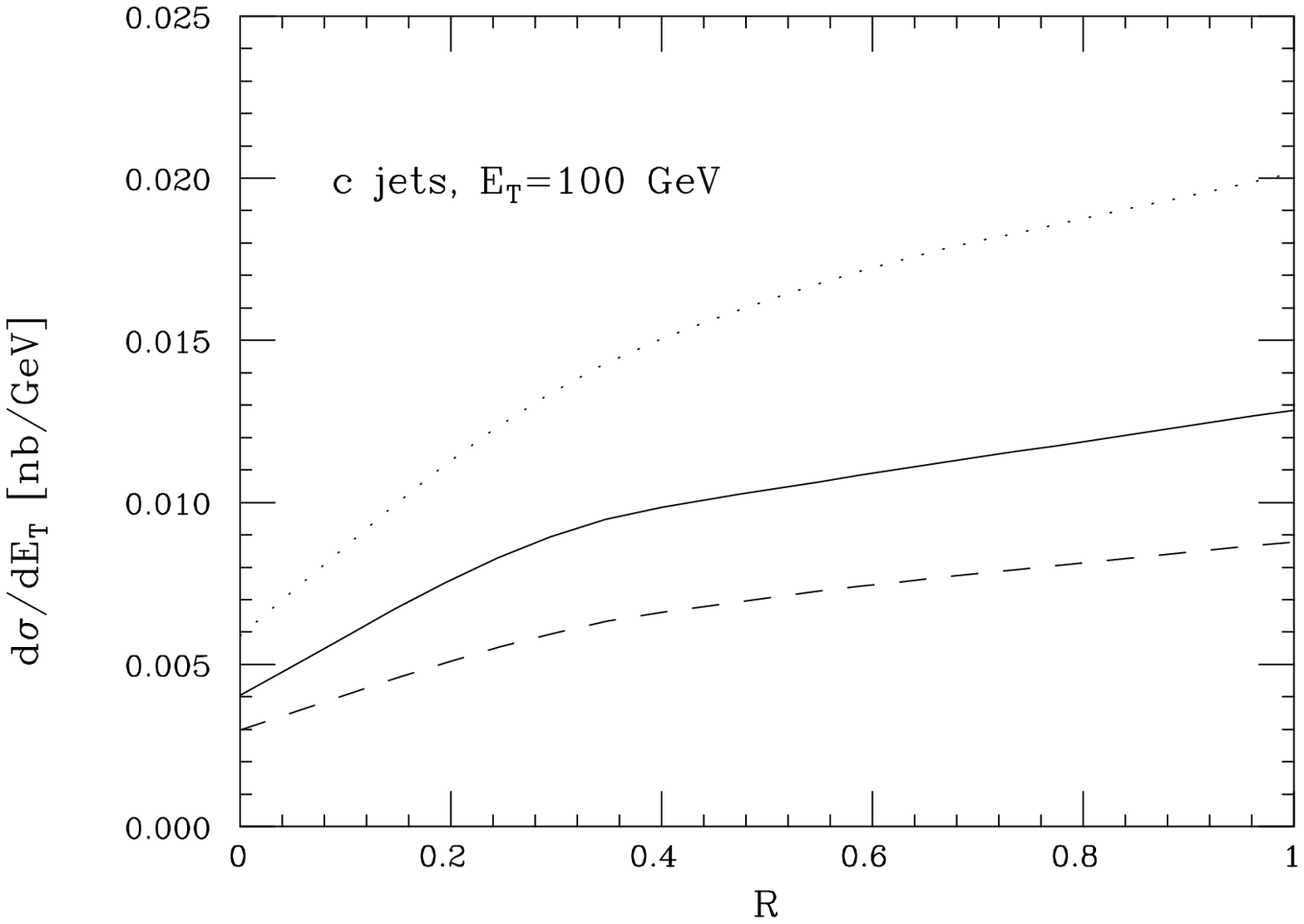,width=0.5\textwidth,clip=}}
\ccaption{}{ \label{fccone}                      
$c$-jet inclusive \et\ rate, as a function of the cone size $R$, for 
\et~=~50~GeV (left) and \et~=~100~GeV (right).}
\end{figure}                                                              
The absolute rates at \et~=~50 and 100~GeV as a                     
function of the cone size are given in fig.~\ref{fbcone}. As explained in
the Appendix, the cross section at $R=0$ is well defined, 
and it is equal to the
open-quark cross section. This should be contrasted with the case of generic
jets, in which the cross section at $R=0$ is not well defined, being
negative at any fixed order in perturbation theory~[\ref{nlojets}].

Similar results for $c$-jets are shown in figs.~\ref{fcfrac} and \ref{fccone}.

The strong scale dependence exhibited by the absolute rates is of the same
size as the one present in the inclusive \pt\ distribution of 
open heavy quarks.
This scale dependence is usually attributed to the 
importance of the gluon splitting contribution. 
This process appears for the first time at \oacube\ and 
is therefore, strictly speaking, calculated with leading-order
accuracy only. One expects therefore that in a regime in which the gluon
splitting contribution is suppressed by the dynamics the scale dependence
should be milder. We will show later that such a suppression takes place for
high-energy heavy-quark jets. Figure~\ref{muratio} shows the scale dependence
of the $b$-jet cross section as a function of \et, for values up to 450~GeV.
\begin{figure}
\centerline{\epsfig{figure=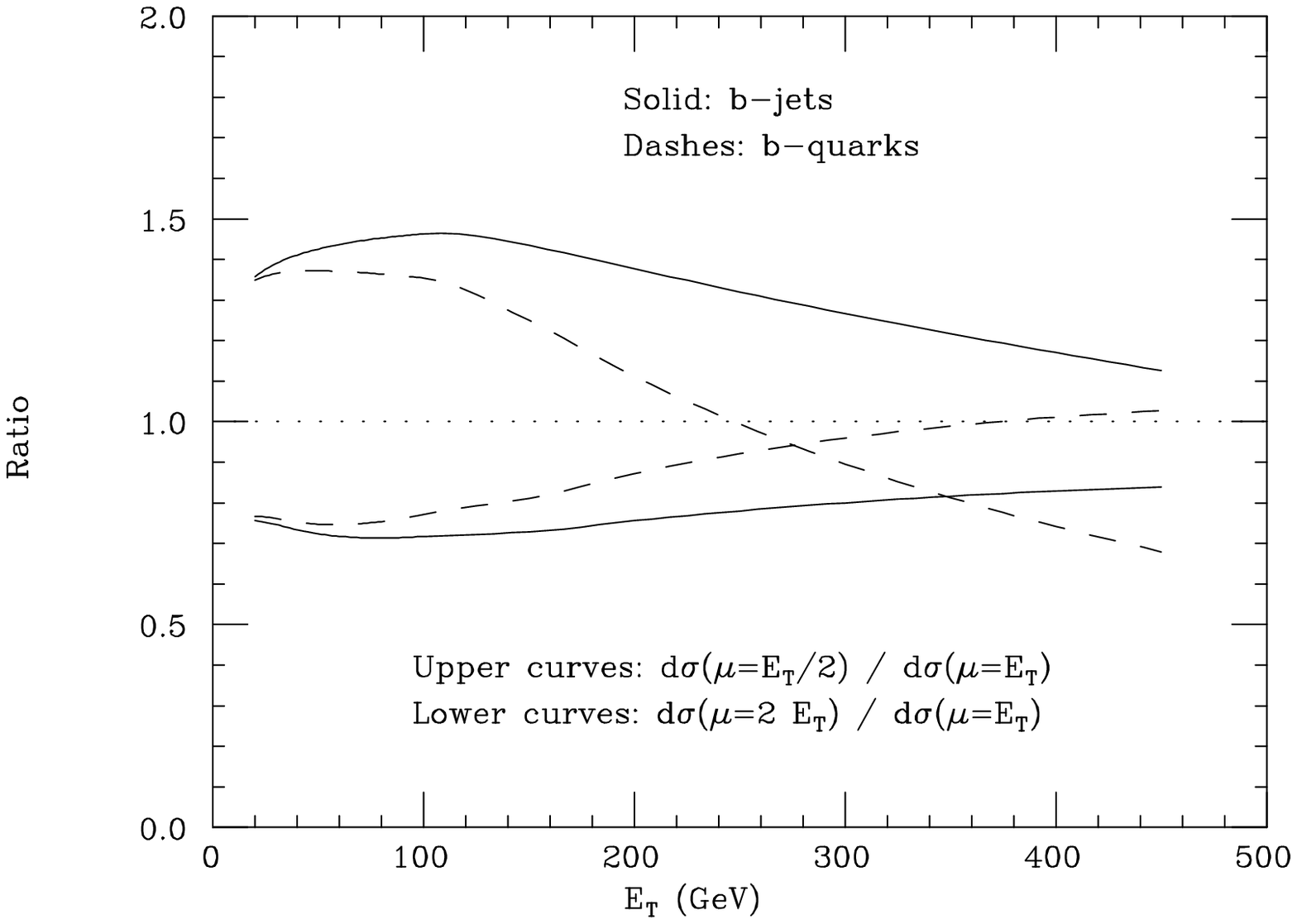,width=0.7\textwidth,clip=}}
\ccaption{}{ \label{muratio}                                      
Scale dependence of the $b$-jet \et\ distribution ($R=0.4$, solid lines) and 
of the open-quark inclusive \et\ distribution (dashed lines). }
\end{figure}                                                              
In the high-$\et$ region the scale dependence is indeed reduced to the
value of $20$\% when the scale is varied in the range $\muo/2<\mu<2\muo$, a
result consistent with the limited scale dependence of the NLO
inclusive-jet cross sections~[\ref{nlojets}].

In spite of the strong scale sensitivity at the smaller values of \et, 
it is reasonable to expect that the ratio of the
$b$- and $c$-jet rates be a stable quantity under scale variations. That
this is indeed the case is shown in fig.~\ref{btoc}. 

\begin{figure}
\centerline{\epsfig{figure=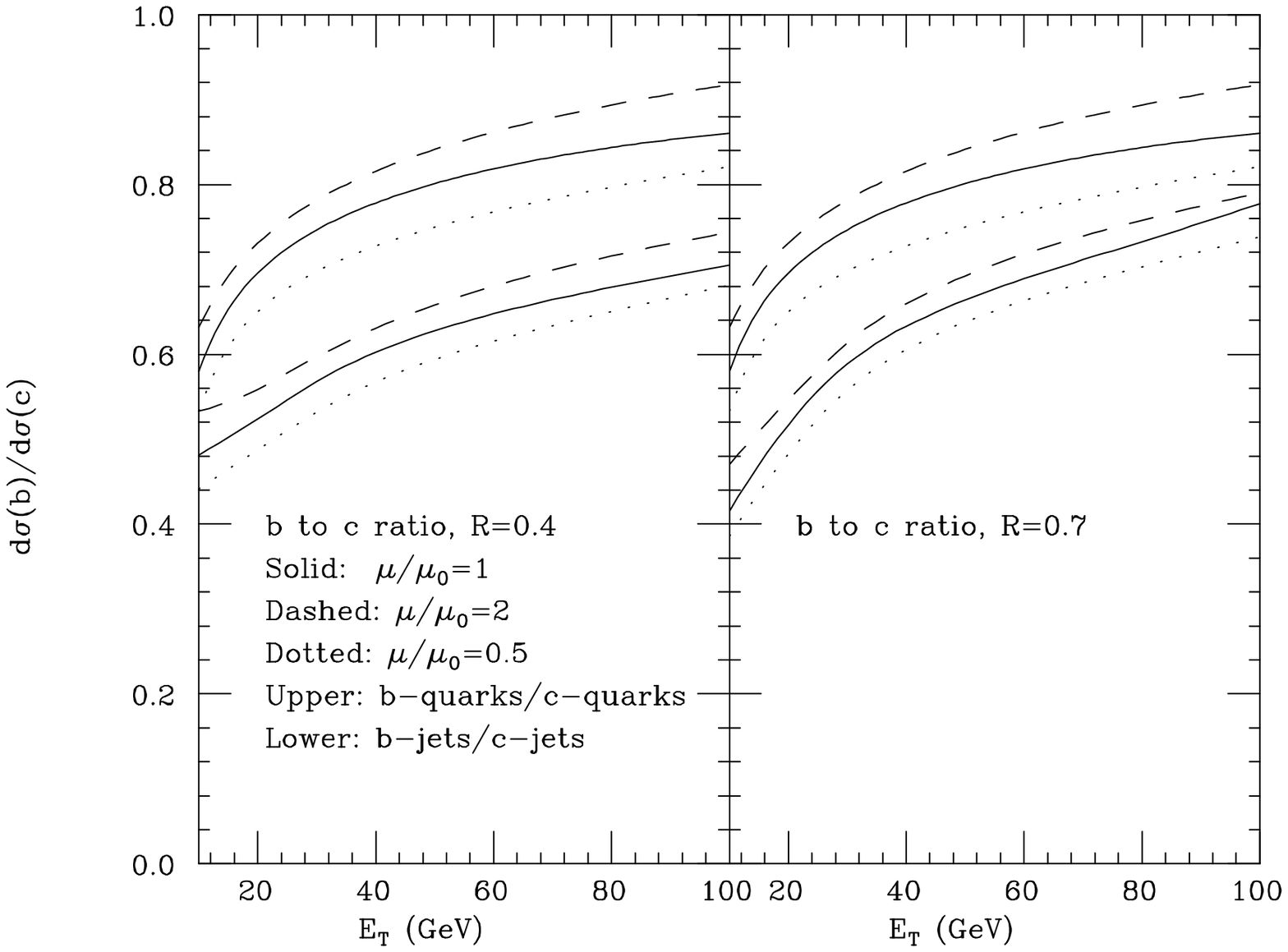,width=0.7\textwidth,clip=}}
\ccaption{}{ \label{btoc}                                      
Ratio of the $b$-jet to $c$-jet \et\ distributions, for different values of the
renormalization and factorization scales ($\mur=\muf=\mu$) for $R=0.4$ (left)
and $R=0.7$ (right). For comparison we also show the ratio of open quark
inclusive \et\ distributions of $b$- and $c$-quarks.}
\end{figure}                                                              

Of direct interest for studies of heavy-flavour tagging and for searches of
possible new physics is the fraction of heavy-quark jets relative to generic
jets. This is also in principle the most straightforward measurement from the
experimental point of view.                      
We present in fig.~\ref{bjetfrac} the ratio of the $b$-jet and inclusive-jet
\et\ distributions~[\ref{nlojets}] (a similar plot for the 
$c$-jet fraction is shown in fig.~\ref{cjetfrac}).
The inclusive-jet \et\ cross section used here  
was calculated with the JETRAD           
program~[\ref{jetrad}], using the same choices of parton densities and
(\mur,\muf) as were adopted for the $b$-jet and $c$-jet calculations.
Contrary to the figures presented so far, which showed results for the heavy
quark only (\ie\ no antiquark contribution), we adopt for this one the
prescription used in the definition of the data presented by CDF~[\ref{koehn}].
The $b$-jets are defined there as jets containing either a $b$ or a $\bar b$
quark, jets containing both being counted only once. We will call these
$b(\bar b)$-jets. This distribution can be obtained by subtracting the
contribution of \bbbar-jets from twice the total $b$-jet rate. 
                                                               
As shown in fig.~\ref{bjetfrac}, the normalization of the heavy-quark jet
fraction depends on the choice of scale as well as on the jet definition. 
\begin{figure}
\centerline{\epsfig{figure=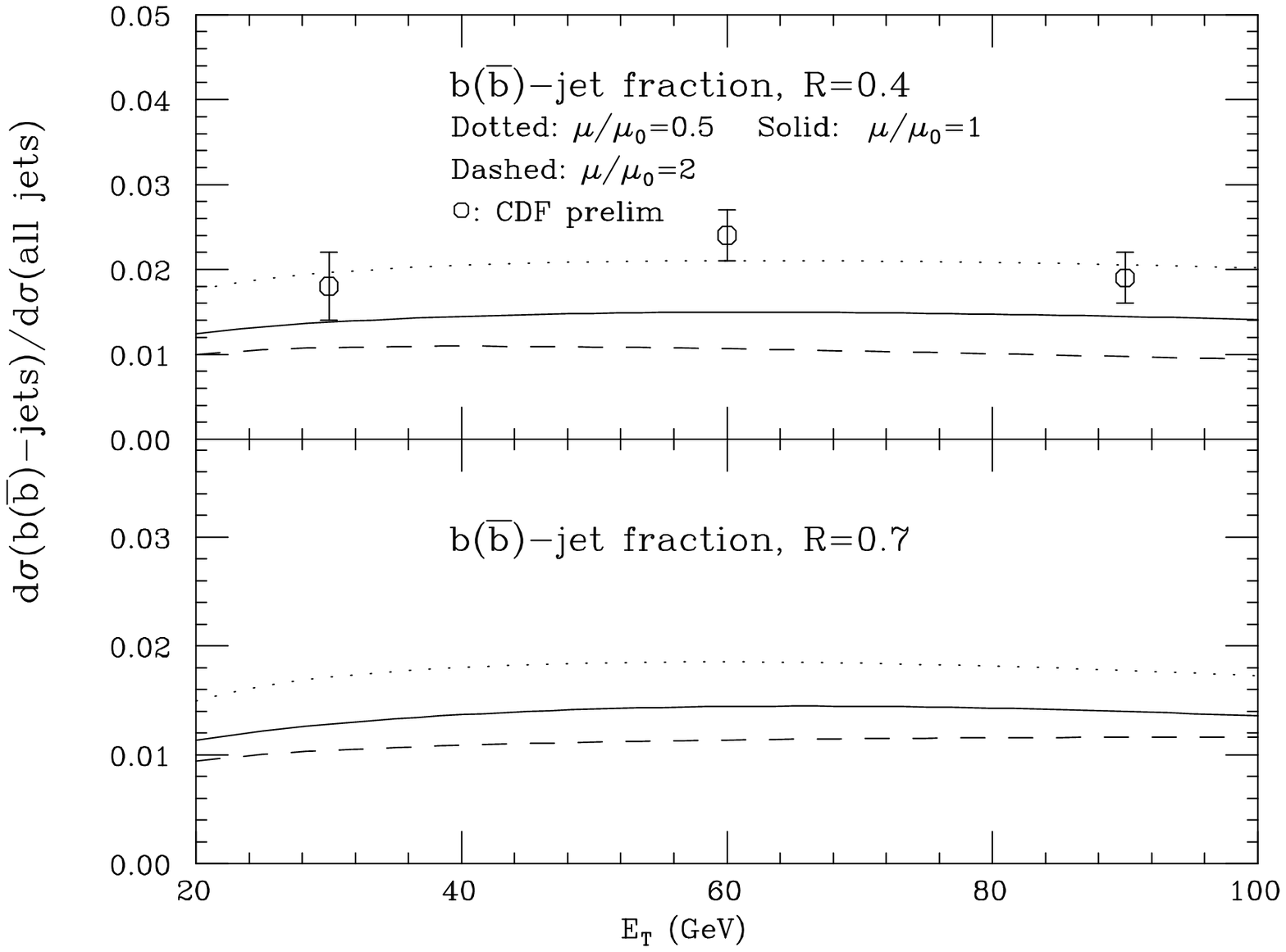,width=0.7\textwidth,clip=}}
\ccaption{}{ \label{bjetfrac}              
Ratio of the $b(\bar b)$-jet to inclusive-jet \et\ 
distributions, for different choices of       
renormalization and factorization scales ($\mur=\muf=\mu$) 
for $R=0.4$ (top) and
$R=0.7$ (bottom). The data points for $R=0.4$ represent preliminary results
from the CDF experiment~[\ref{koehn}],  for which only the statistical
uncertainty is shown.}
\end{figure}                                                              
In particular, the ambiguity induced by the change of scale
is of ${\cal O}(2)$. While this            
uncertainty prevents an accurate prediction of the heavy-quark jet fractions,
it is important to point out that the choice of scale for this process is not
independent of the scale chosen to predict the inclusive open-quark \pt\
distributions. Since the data  on both the                            
bottom quark~[\ref{bdata},\ref{fmnr}] and the inclusive-jet \et\
spectra~[\ref{jetdata},\ref{jetdataD0}]
strongly support the choice $\mur$, $\muf \sim
\muo/2$, or smaller, we suggest that this is the 
best choice for the scales to be
used in the prediction of the $b$-jet fraction.

It is interesting to notice that with this choice of scale there is good
agreement between the theoretical prediction and the CDF data, at least in the
case of $b$-jets. Notice also that, while the data on inclusive $b$-hadron
distributions require the choice of even smaller scales (of the order of
$\muo/4$~[\ref{fmnr}]), the measurement of the $b(\bar b)$-jet rates indicates
that  the choice $\mu=\muo/2$ is adequate. 
As for the large disagreement with the charm data, we have no significant 
comment to make. Hopefully, additional data will soon be available, as well as
estimates of the experimental systematics. Should the disagreement persist,
this would indicate the presence of theoretical systematics not accounted for
by the standard procedure of exploring the scale dependence of the rates.
Notice that the largest contribution expected from higher-order perturbative
corrections is given by the production of $c\bar c$ pairs from softer gluons
emitted during the gluon shower evolution. However these effects have been
estimated in ref.~[\ref{hvqjet}], and have been shown to be negligible at the
energies of interest for the current measurements.

\begin{figure}
\centerline{\epsfig{figure=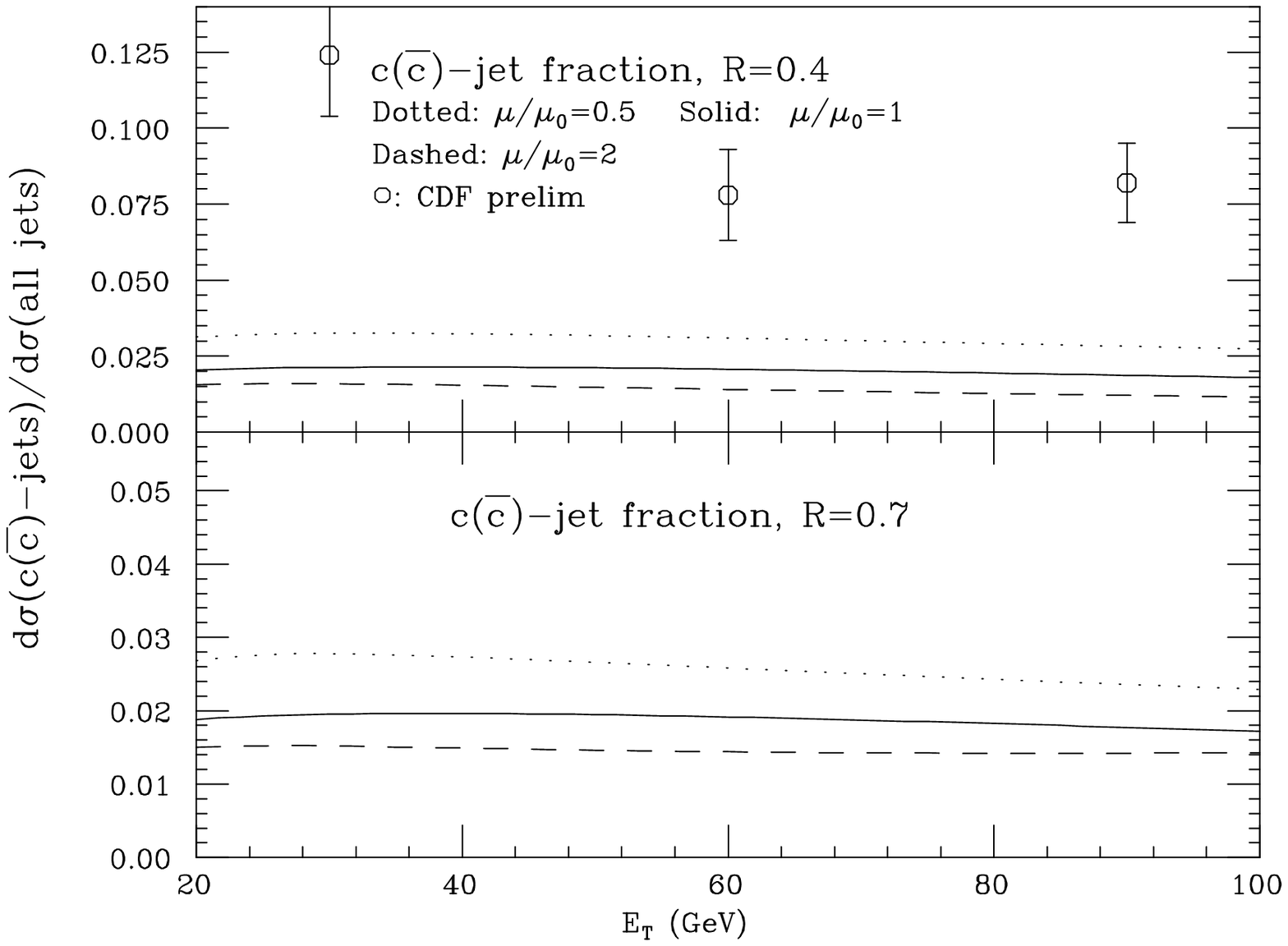,width=0.7\textwidth,clip=}}
\ccaption{}{ \label{cjetfrac}       
Ratio of the $c(\bar c)$-jet to inclusive-jet \et\ 
distributions, for different choices of 
renormalization and factorization scales (\mur=\muf=$\mu$) 
for $R=0.4$ (top) and
$R=0.7$ (bottom). The data points for $R=0.4$ represent preliminary results
from the CDF experiment~[\ref{koehn}],  for which only the statistical
uncertainty is shown.}
\end{figure}                                                              

To conclude this section, we discuss the behaviour of the $b$-jet production
cross section at high \et. 
This item is interesting in view of the 
discrepancy reported by CDF~[\ref{jetdata}] in the
tail of the jet distribution. If this discrepancy could not be accommodated by
new theoretical developments in QCD~[\ref{cmnt}] or in the fitting of parton
densities~[\ref{cdfpdf}], a study of the flavour composition of these 
high-energy jets could help in understanding the nature of the phenomenon. 

\begin{figure}
\centerline{\epsfig{figure=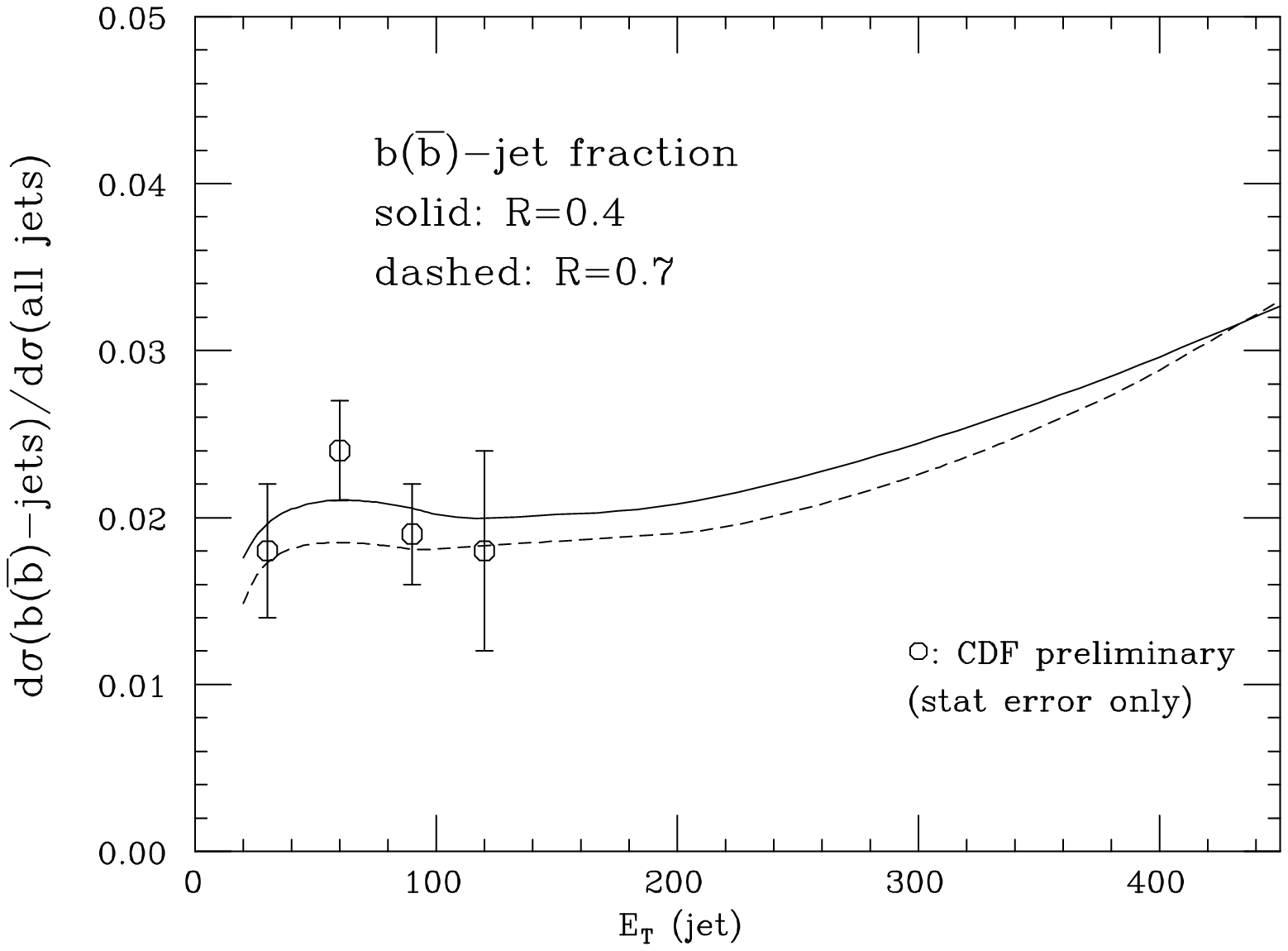,width=0.7\textwidth,clip=}}
\ccaption{}{ \label{highet}          
Ratio of the $b(\bar b)$-jet to inclusive-jet \et\ 
distributions for $\mur=\muf=\mu/2$ and with $R=0.4$ (solid) or
$R=0.7$ (dashes). The data points
are preliminary CDF data~[\ref{koehn}], obtained with $R=0.4$.}
\end{figure}                                                              

Figure~\ref{highet} shows the $b(\bar b)$-jet fraction for \et\ values up
to 450 GeV, for two different values of the cone size ($R=0.4$ and $R=0.7$)
and at $\mur=\muf=\muo/2$. 
Notice that while the fraction remains constant
through most of the \et\ range, a rise is observed above 300~GeV. To better
understand the origin of this rise, we present in fig.~\ref{bhtot}a the
separate contribution to the $b$-jet cross section                    
of the three possible initial states, $gg$, $q\bar q$ and $qg$. Notice that the
$q\bar q$ contribution becomes dominant for $\et>250$~GeV.
Figures~\ref{bhtot}b--d show, for each individual channel,  the separate
contribution of the open-quark and \bbbar-jet components. For \et\ large
enough, the dominant component of the $gg$ and $qg$ channels is given by the
\bbbar-jet contribution, because of the gluon-splitting dominance. In the case
of the $q\bar q$ channel, on the contrary, the \bbbar-jet term is always
suppressed, and most of the $b$-jets are composed of a single $b$
quark, often accompanied by a nearby gluon. 
We conclude that at high \et\ the dominant mechanism for the production of a
$b$-jet is the $s$-channel annihilation of light quarks. Since at high \et\
mass effects are negligible, 
1/5 of the jets produced in $s$-channel annihilation are $b$-jets. A simple
LO calculation shows that the fraction of the two-jet rate due to $s$-channel
light-quark annihilation is about 20\% at \et~=~450~GeV, giving an overall
$b$-jet over inclusive-jet fraction of approximately 4\%. This explains the
rise of the $b$-fraction at high \et, and provides a nice consistency check of
our results. Notice also that while the probability that a gluon-jet 
will split into a \bbbar\ pair grows at large \et\ 
faster that what predicted by the  \oacube\
calculation~[\ref{hvqjet}]\footnote{This happens because  of pairs emitted at
higher-order from the gluons of the shower.},  the fraction of primary gluons
in the final state is so small that the overall effect on our predictions is
negligible. 

\begin{figure}
\centerline{\epsfig{figure=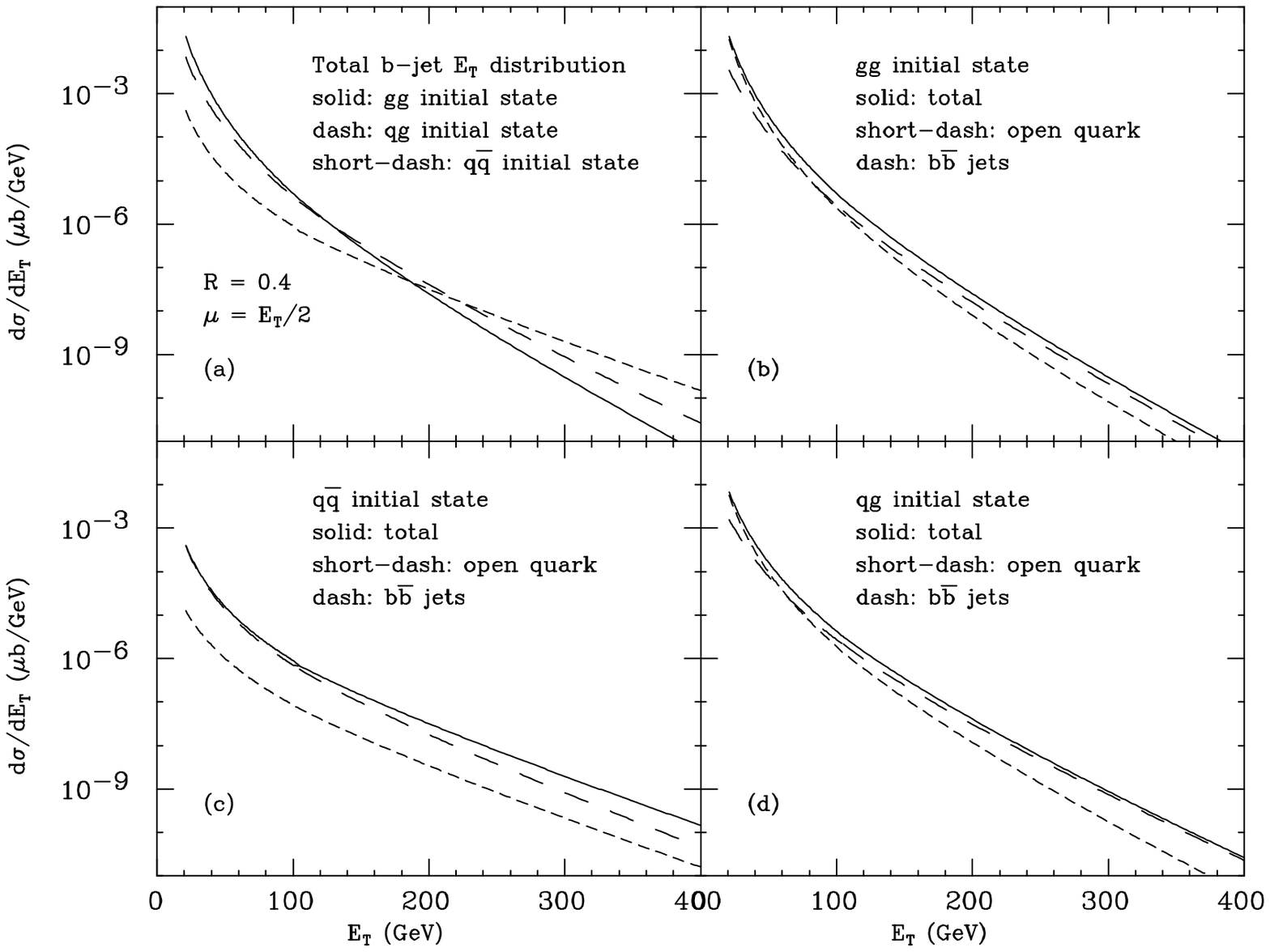,width=\textwidth,clip=}}
\ccaption{}{ \label{bhtot}  
Initial state composition of the $b$-jet production processes,
calculated for $\mur=\muf=\mu/2$ and $R=0.4$ (upper left).
Different components of the                             
production processes: $gg\;\to\; b$-jet (upper right), $q{\bar q} \;\to\;
b$-jet (lower left) and $qg\;\to\; b$-jet (lower right).}
\end{figure}                                                              
            
\section{Conclusions}
We presented in this paper a calculation of the production of jets containing
heavy quarks, at NLO in perturbative QCD. 
The techniques we employed represent a
further elaboration of standard methods developed in recent years for the study
of higher-order processes in hadronic collisions. As a phenomenological
application of our results, we presented a detailed discussion of the rates and
properties of charm and bottom jets produced at the Tevatron $p\bar p$
collider. We found that some distributions, such as the ratio of bottom to
charm jets as a function of the jet transverse energy, are rather independent
of the choice of renormalization and factorization scales. We presented results
for the $b$- and $c$-jet fraction of inclusive jets, and discussed in detail
the properties and composition of very high \et\ jets. We found that these
vary significantly across the \et\ range measurable at the Tevatron. We
provided predictions for the $b$-jet fractions at high \et, to be compared
with what could be measured by the Tevatron collider experiments.
\\[0.3cm]                                                                   
\section*{Acknowledgements}
One of the authors (S.F.) wishes to thank the members of the
FNAL theory division,
where part of this work was performed, for their kind hospitality. The authors
thank Phil Koehn, Arthur Maciel and Andre Sznajder for providing them with 
useful information on the experimental measurements, and Walter Giele and
Nigel Glover for providing them with a copy of the JETRAD code.     
\appendix
\section{Heavy-quark jet definition}
In this appendix we report the technical details that were
not explicitly given in the previous sections.       
In the following, we will always indicate the kinematics
of a partonic process as 
\beq
a(p_1)+b(p_2)\,\rightarrow\,Q(k_1)+\Qb(k_2)+c(k),
\eeq
where $c$ is the final state light parton present
in the two-to-three processes.

We begin by considering the leading-order cross section, 
eq.~(\ref{LOxsec}). As discussed in section 2, in this
case the heavy-quark jet coincides with the heavy quark.
The $\Stwo$ measurement function formally states
this obvious fact
\beq
\Stwo=\delta(\EJT-E_{\sss 1T})\delta(\eta_\Js-\eta_{\sss 1})
\delta^{1-2\ep}(\varphi_\Js-\varphi_{\sss 1}).
\label{S2def}
\eeq
Here we denote by $\EJT$ the transverse energy of
the heavy-quark jet, and with $\eta_\Js$ and $\varphi_\Js$ its
pseudorapidity and azimuthal angle respectively. 
$E_{\sss 1T}=k_{\sss 1}^0/\cosh(\eta_{\sss 1})$ is the transverse 
energy of the heavy quark; notice that the transverse energy is 
equal to the transverse momentum only in the case of massless particles.

We then turn to the next-to-leading order contribution, 
eq.~(\ref{NLOxsec}). Due to eq.~(\ref{S2def}), the heavy-quark jet
cross section will be different from the open-heavy-quark one
only in the real part, eq.~(\ref{realxsec}).
Using the merging algorithm of ref.~[\ref{snowmass}], we have
\beqn
\Sthree&=&\delta(\EJT-E_{\sss 1T})\delta(\eta_\Js-\eta_{\sss 1})
\delta^{1-2\ep}(\varphi_\Js-\varphi_{\sss 1})
\nonumber \\*&&\phantom{xxxx}
\times\theta\Bigg(\abs{\omega_{\sss 1}-\omega}>
g(E_{\sss 1T},E_{\sss T})\Bigg)
\theta\Bigg(\abs{\omega_{\sss 1}-\omega_{\sss 2}}>
g(E_{\sss 1T},E_{\sss 2T})\Bigg)
\nonumber \\*
&+&\delta(\EJT-E_{\sss 1T}-E_{\sss T})
\delta\left(\eta_\Js-\frac{\eta_{\sss 1} E_{\sss 1T}
+\eta E_{\sss T}}{\EJT}\right)
\delta^{1-2\ep}\left(\varphi_\Js
-\frac{\varphi_{\sss 1} E_{\sss 1T}+\varphi E_{\sss T}}{\EJT}\right)
\nonumber \\*&&\phantom{xxxx}
\times\theta\Bigg(\abs{\omega_{\sss 1}-\omega}<g(E_{\sss 1T},E_{\sss T})\Bigg)
\nonumber \\*
&+&\delta(\EJT-E_{\sss 1T}-E_{\sss 2T})
\delta\left(\eta_\Js-\frac{\eta_{\sss 1} E_{\sss 1T}
+\eta_{\sss 2} E_{\sss 2T}}{\EJT}\right)
\delta^{1-2\ep}\left(\varphi_\Js
-\frac{\varphi_{\sss 1} E_{\sss 1T}
+\varphi_{\sss 2} E_{\sss 2T}}{\EJT}\right)
\nonumber \\*&&\phantom{xxxx}
\times\theta\Bigg(\abs{\omega_{\sss 1}-\omega_{\sss 2}}
<g(E_{\sss 1T},E_{\sss 2T})\Bigg),
\label{S3def}
\eeqn
where
\beq
g(x,y)=\frac{x+y}{max(x,y)}R
\label{gfundef}
\eeq
and
\beq
\abs{\omega_i-\omega_j}=\sqrt{(\eta_i-\eta_j)^2+(\varphi_i-\varphi_j)^2}\,.
\eeq
In eq.~(\ref{gfundef}), $R$ is the usual jet-resolution parameter,
which defines the cone size in the pseudorapidity-azimuthal angle plane.
The quantities appearing in previous equations are expressed
in the laboratory frame.

It is well known that perturbatively calculated QCD cross sections
have a divergent behaviour even after the ultraviolet renormalization,
in those regions in which a massless parton is soft or collinear to
another massless parton. These singularities cancel after the
virtual part, the real part and the collinear counterterms are
added together. To perform this cancellation analytically is therefore
important, in order to disentangle the various singular contributions to
the cross section. To this end, the preliminary step in our formalism
is to study the behaviour of the measurements functions in the
soft and collinear regions. In the case at hand, the only non-trivial
case is the one of the $\Sthree$ function, since the two-to-three
processes are the only ones in which a massless parton is emitted.
It is straightforward to see that the following equations hold
\beqn
\lim_{k^0\to 0}\Sthree&=&\Stwo\,,
\label{S3lim1}
\\
\lim_{\vec{k}\to\vec{k}_1}\Sthree
=\lim_{\vec{k}\to\vec{k}_2}\Sthree&=&\Stwo\,.
\label{S3lim2}
\eeqn
This properties guarantee the infra-red safeness of the heavy-quark jet 
cross section. Notice that, due to the fact that the heavy quark
is massive, we do not have to care about the final state collinear
emission, which is relevant, on the other hand, for the generic-jet
cross section calculations.                     

We now have to disentangle the singularities appearing
in the heavy-quark jet cross section. For this purpose, we use the same
method as was recently presented in ref.~[\ref{FKS}]. We parametrize
the four-momentum of the outgoing light parton as
\beq
k=\frac{\sqrt{S}}{2}\xi(1,\sqrt{1-y^2}\,\vec{e}_{\sss T},y)\,,
\label{kpar}
\eeq
where $\vec{e}_{\sss T}$ is a $(2-2\ep)$-dimensional unitary vector in the
transverse momentum space, $-1\le y\le 1$ and $0\le\xi\le\xi_{\sss L}$; the
collinear and soft limits are $y\to\pm 1$ and $\xi\to 0$ respectively.
With this parametrization, we have
\beq
E_{\sss T}=\frac{\sqrt{S}}{2}\xi\sqrt{1-y^2}\,.
\label{kTmod}
\eeq
Defining
\beq
d\Phi^{(n)}(k)=\frac{d^{n-1}k}{(2\pi)^{n-1}2k^0}\,,
\eeq
we get
\beq
E_{\sss T}^{-2}d\Phi^{(4-2\ep)}(k)
=\frac{1}{2(2\pi)^{3-2\ep}}\left(\frac{\sqrt{S}}{2}
\right)^{-2\ep}\xi^{-1-2\ep}\left(1-y^2\right)^{-1-\ep}
d\xi dy d\Omega^{(2-2\ep)}\,,
\label{dphiglu}
\eeq
where $d\Omega^{(2-2\ep)}$ is the angular measure in $2-2\ep$ dimensions.
This form is suitable for disentangling the singular contributions
to the real cross section. We use the identity
\beq
\xi^{-1-2\ep}\left(1-y^2\right)^{-1-\ep}={\cal D}(\xi,y)
+{\cal R}(\xi,y)+{\cal O}(\ep),
\label{distrid}
\eeq
where
\beqn
{\cal D}(\xi,y)&=&-\frac{\xi_{cut}^{-2\ep}}{2\ep}
\delta(\xi)\left(1-y^2\right)^{-1-\ep}
\nonumber \\*
&-&\frac{(2y_{cut})^{-\ep}}{2\ep}
\left[\delta(1-y)+\delta(1+y)\right]\left[\left(\frac{1}{\xi}\right)_c
-2\ep\left(\frac{\log\xi}{\xi}\right)_c\right],
\label{Ddef}
\\
{\cal R}(\xi,y)&=&\frac{1}{2}\left(\frac{1}{\xi}\right)_c
\left[\left(\frac{1}{1-y}\right)_c
+\left(\frac{1}{1+y}\right)_c\right].
\label{Rdef}
\eeqn
The distributions in eqs.~(\ref{Ddef}) and~(\ref{Rdef}) are
defined as follows
\beqn   
\Bigg\langle \left(\frac{1}{\xi}\right)_c,f \Bigg\rangle &=&\int_0^1 d\xi\,
\frac{f(\xi)-f(0)\Th(\xi_{cut}-\xi)}{\xi}\,,             
\label{uoxidef}
\\
\Bigg\langle \left(\frac{1}{1-y}\right)_c,f \Bigg\rangle &=&\int_{-1}^{1}dy\,
\frac{f(y)-f(1)\Th(y-1+y_{cut})}{1-y}\,,                 
\\
\Bigg\langle \left(\frac{1}{1+y}\right)_c,f \Bigg\rangle &=&\int_{-1}^{1}dy\,
\frac{f(y)-f(-1)\Th(-y-1+y_{cut})}{1+y}\,,
\eeqn
where $0<\xi_{cut}\leq 1$ and $0<y_{cut}\leq 2$ are arbitrary parameters,
which can be freely chosen to improve the convergence of the results     
in the numerical computations. Writing the three-body phase space as
\beq
d\Phi_3=E_{\sss T}^2\,d\tilde{\Phi}_2\,E_{\sss T}^{-2}\,d\Phi^{(4-2\ep)}(k) \;
,
\label{Phi3id}
\eeq
(by construction, $d\tilde{\Phi}_2$ is basically identical to the
two-body phase space for the $Q\bar{Q}$ pair introduced in
eq.~(\ref{LOxsec}), except for the delta over the four-momentum,
which in the present case reads $\delta(p_1+p_2-k_1-k_2-k)$),
we can exploit eqs.~(\ref{dphiglu}) and~(\ref{distrid}) to write
eq.~(\ref{realxsec}) as follows (we neglect ${\cal O}(\ep)$ terms)
\beqn
d\sigma^{(r)}_{ab}&=&\Sthree {\cal M}_{ab}^{(r)}\,E_{\sss T}^2\, 
d\tilde{\Phi}_2\,\frac{1}{2(2\pi)^{3-2\ep}}
\left(\frac{\sqrt{S}}{2}\right)^{-2\ep}
{\cal D}(\xi,y)d\xi dy d\Omega^{(2-2\ep)} d\mu_\Js
\nonumber \\*&+&
\Sthree {\cal M}_{ab}^{(r)}\,E_{\sss T}^2\,d\tilde{\Phi}_2\,
\frac{1}{2(2\pi)^{3-2\ep}}\left(\frac{\sqrt{S}}{2}\right)^{-2\ep}
{\cal R}(\xi,y)d\xi dy d\Omega^{(2-2\ep)}d\mu_\Js\,.\phantom{aaaa}
\label{sigmajet2}
\eeqn
The measurement function $\Sthree$ in the first term on the RHS
of eq.~(\ref{sigmajet2}) can be substituted with $\Stwo$.
In fact, the $\delta(\xi)$ and $\delta(1\pm y)$ contained
in the factor ${\cal D}(\xi,y)$ allow us to take the soft and 
collinear limits of that term, that is, to exploit 
eqs.~(\ref{S3lim1}) and~(\ref{S3lim2}). We can also write
\beq
\Sthree=\Stwo+\Stilde
\label{Sthreesplit}
\eeq
where
\beqn
\Stilde&=&\delta(\EJT-E_{\sss 1T})\delta(\eta_\Js-\eta_{\sss 1})
\delta^{1-2\ep}(\varphi_\Js-\varphi_{\sss 1})
\nonumber \\*&&\phantom{xxxx}
\times\Bigg[-\theta\Bigg(\abs{\omega_{\sss 1}-\omega}<
g(E_{\sss 1T},E_{\sss T})\Bigg)
-\theta\Bigg(\abs{\omega_{\sss 1}-\omega_{\sss 2}}<
g(E_{\sss 1T},E_{\sss 2T})\Bigg)\Bigg]
\nonumber \\*
&+&\delta(\EJT-E_{\sss 1T}-E_{\sss T})
\delta\left(\eta_\Js-\frac{\eta_{\sss 1} E_{\sss 1T}
+\eta E_{\sss T}}{\EJT}\right)
\delta^{1-2\ep}\left(\varphi_\Js
-\frac{\varphi_{\sss 1} E_{\sss 1T}+\varphi E_{\sss T}}{\EJT}\right)
\nonumber \\*&&\phantom{xxxx}
\times\theta\Bigg(\abs{\omega_{\sss 1}-\omega}<g(E_{\sss 1T},E_{\sss T})\Bigg)
\nonumber \\*
&+&\delta(\EJT-E_{\sss 1T}-E_{\sss 2T})
\delta\left(\eta_\Js-\frac{\eta_{\sss 1} E_{\sss 1T}
+\eta_{\sss 2} E_{\sss 2T}}{\EJT}\right)
\delta^{1-2\ep}\left(\varphi_\Js
-\frac{\varphi_{\sss 1} E_{\sss 1T}+\varphi_{\sss 2} E_{\sss 2T}}{\EJT}\right)
\nonumber \\*&&\phantom{xxxx}
\times\theta\Bigg(\abs{\omega_{\sss 1}-\omega_{\sss 2}}
<g(E_{\sss 1T},E_{\sss 2T})\Bigg),
\label{S3tilde}
\eeqn
having used the identity
\beq
\theta(a>b)\theta(c>d)=1-\theta(a<b)-\theta(c<d)
+\theta(a<b)\theta(c<d)
\label{thetaid}
\eeq
in the first term on the RHS of eq.~(\ref{S3def}). Notice that the 
first term on the RHS of eq.~(\ref{S3tilde}) should also have a 
contribution of the kind
\beq
\theta\Bigg(\abs{\omega_{\sss 1}-\omega}<g(E_{\sss 1T},E_{\sss T})\Bigg)
\theta\Bigg(\abs{\omega_{\sss 1}-\omega_{\sss 2}}
<g(E_{\sss 1T},E_{\sss 2T})\Bigg).
\eeq
Nevertheless, this quantity is always equal to zero, since otherwise
it would be possible to merge in a jet all three particles in
the final state. Equation~(\ref{sigmajet2}) therefore becomes 
\beqn
d\sigma^{(r)}_{ab}&=&\Stwo {\cal M}_{ab}^{(r)}\,E_{\sss T}^2\, 
d\tilde{\Phi}_2\,\frac{1}{2(2\pi)^{3-2\ep}}
\left(\frac{\sqrt{S}}{2}\right)^{-2\ep}
{\cal D}(\xi,y)d\xi dy d\Omega^{(2-2\ep)} d\mu_\Js
\nonumber \\*&+&
\Stwo {\cal M}_{ab}^{(r)}\,E_{\sss T}^2\,d\tilde{\Phi}_2\,
\frac{1}{2(2\pi)^{3-2\ep}}\left(\frac{\sqrt{S}}{2}\right)^{-2\ep}
{\cal R}(\xi,y)d\xi dy d\Omega^{(2-2\ep)} d\mu_\Js
\nonumber \\*&+&
\Stilde {\cal M}_{ab}^{(r)}\,E_{\sss T}^2\,d\tilde{\Phi}_2\,
\frac{1}{2(2\pi)^{3-2\ep}}\left(\frac{\sqrt{S}}{2}\right)^{-2\ep}
{\cal R}(\xi,y)d\xi dy d\Omega^{(2-2\ep)} d\mu_\Js\,.\phantom{aaa}
\label{sigmajet3}
\eeqn
Using again eqs.~(\ref{distrid}) and~(\ref{Phi3id}) we finally get
\beq
d\sigma^{(r)}_{ab}=d\sigma^{(r,open)}_{ab}+d\Delta_{ab}\,,
\label{sigmajetfinal}
\eeq
with
\beqn
d\sigma^{(r,open)}_{ab}&=&\Stwo {\cal M}_{ab}^{(r)}\,d\Phi_3\,d\mu_\Js\,,
\label{sigmareal}
\\
d\Delta_{ab}&=&\Stilde {\cal M}_{ab}^{(r)}
E_{\sss T}^2\,d\tilde{\Phi}_2\,\frac{1}{2(2\pi)^{3-2\ep}}
\left(\frac{\sqrt{S}}{2}\right)^{-2\ep}
{\cal R}(\xi,y)d\xi dy d\Omega^{(2-2\ep)}d\mu_\Js\,.\phantom{aaa}
\label{Delta}
\eeqn
Equation~(\ref{sigmareal}) is just the real contribution to 
open-heavy-quark production. We have therefore
\beq
d\sigma^{(open)}_{ab}=d\sigma^{(0)}_{ab}+d\sigma^{(v)}_{ab}+
d\sigma^{(r,open)}_{ab}\,,
\label{sigmaopen}
\eeq
where the first two quantities on the RHS of this equation were
given in eqs.~(\ref{LOxsec}) and~(\ref{virtxsec}), respectively.

We observe that, although eq.~(\ref{Sthreesplit}) has a general
validity, in the case of generic-jet production it is totally useless,
since eq.~(\ref{Delta}) would display divergences due to final
state collinear emission. These divergences would be cancelled
by corresponding ones in eq.~(\ref{sigmareal}). On the
other hand, in the heavy-quark jet case the mass of the heavy quark is acting
as a cutoff against these divergences, and eq.~(\ref{Delta}) is 
finite and can be numerically integrated. In this sense,
eq.~(\ref{Delta}) is peculiar of the heavy-quark jet cross section.
Notice finally that this term vanishes when $R \; \to \; 0$, that is, 
when no merging is performed; this is physically sensible, since
in the absence of merging we expect the heavy-quark jet cross section to
coincide with the open-heavy-quark one, as is formally
stated by eq.~(\ref{sigmajetfinal}).       

To integrate numerically eq.~(\ref{Delta}),
we observe that in the collinear limits $y \; \to\; \pm 1$ and in the soft
limit $\xi\to 0$ the quantity $\Stilde$ vanishes. It follows that 
the subtractions at $y=\pm 1$ and $\xi=0$, implicitly contained in the 
factor ${\cal R}(\xi,y)$, eq.~(\ref{Rdef}), are actually immaterial, 
and we can therefore perform in eq.~(\ref{Delta}) the formal substitutions
\beqn
\left(\frac{1}{1\pm y}\right)_c &\rightarrow&\frac{1}{1\pm y}\,,
\\
\left(\frac{1}{\xi}\right)_c &\rightarrow&\frac{1}{\xi}\,.
\eeqn
Furthermore, since all divergences have been properly
regulated, we can set $\ep=0$ in eq.~(\ref{Delta}). After
some algebra, we get
\beq
d\Delta_{ab}=\Stilde {\cal M}_{ab}^{(r)}\,d\Phi_3\,\jetmsrf\,.
\label{Deltafinal}
\eeq
Using eqs.~(\ref{sigmaopen}) and~(\ref{Deltafinal}) we get 
eq.~(\ref{xsecsplit}) and the explicit form for $d\Delta_{ab}$. 
We finally observe that, since eq.~(\ref{Deltafinal}) is free of 
divergences, it is suitable for numerical integration.

\begin{reflist}
\item \label{NDE}                               
        P.~Nason, S.~Dawson and R.~K.~Ellis, 
        \np{B303}{88}{607}; {\bf B327}{(1988)}{49};\\
        W.~Beenakker, H.~Kuijf, W.L. van Neerven and  J. Smith, 
        \pr{D40}{89}{54};
        W.~Beenakker et al., \np{B351}{91}{507}.         
\item \label{MNR}                      
 M.~Mangano, P.~Nason and G.~Ridolfi, \np{B373}{92}{295}.
\item \label{ua1}
C. Albajar et al., UA1 Collaboration, \pl{B256}{91}{121}.
\item \label{bdata}
F. Abe et al., CDF Collaboration, \prl{68}{92}{3403}; {\bf 69}(1992)3704;
{\bf 71}(1993)500, 2396, 2537; {\bf 75}(1995)1451; \pr{D53}{96}{1051}\\
S. Abachi et al., D0 Collaboration, \prl{74}{95}{3548}.
\item \label{greco}
M. Cacciari and M. Greco,
        \np{B421}{94}{530}.    
\item \label{CSS} 
 J.~C.~Collins, D.~E.~Soper and G.~Sterman, in {\it Perturbative
 Quantum Chromodynamics}, 1989, ed. Mueller, World Scientific,
 Singapore, and references therein.
\item \label{snowmass}            
 F.~Aversa et al., Proceedings of the Summer Study on 
 High Energy Physics, Research Directions for the Decade, 
 Snowmass, CO, 1990, edited by E. Berger (World Scientific, Singapore, 1991).
\item\label{KS}                                                              
 Z.~Kunszt and D.~E.~Soper, \pr{D46}{92}{192}.
\item\label{FKS}
 S.~Frixione, Z.~Kunszt and A.~Signer, preprint ETH-TH/95-42, \\
 SLAC-PUB-95-7073, hep-ph/9512328, to appear in {\it Nucl. Phys.}~{\bf B}.
\item \label{koehn}
  P. Koehn, for the CDF Collaboration, Proceedings of the 1995 CTEQ "Workshop
  on Collider Physics", Michigan State University.
\item \label{MRSAP}                    
  A.D. Martin, R.G. Roberts and W.J. Stirling, \pr{D50}{94}{6734}.
\item \label{hvqjet}                                              
 A.H. Mueller and P. Nason, \pl{B157}{85}{226}; \np{B266}{86}{265};\\
 M.L. Mangano and P. Nason, \pl{B285}{92}{160};\\
 M.H. Seymour, \np{B436}{95}{163}.                               
\item \label{nlojets}
   F. Aversa, P. Chiappetta, M. Greco and J.P.Guillet,
   \np{B327}{89}{105},  \prl{65}{90}{401};\\
   S. Ellis, Z. Kunszt and D. Soper,
    \pr{D40}{89}{2188}, \prl{64}{90}{2121}.
\item \label{jetrad}
  W.T. Giele, E.W.N. Glover and D.A. Kosower, \prl{73}{94}{2019}.
\item \label{fmnr}                                                         
 S. Frixione,  M.~Mangano, P.~Nason and G.~Ridolfi, \np{B431}{94}{453}.
\item \label{jetdata}                                                
F. Abe et al., CDF Collaboration, \prl{68}{92}{1104}; Fermilab-Pub-96/20-E, 
hep-ex/9601008
\item \label{jetdataD0} 
G. Blazey, for the D0 Collaboration, presented at the Rencontres de Moriond,
QCD and Hadronic Interactions, 24-30 March 1996.
\item \label{cmnt}
   S. Catani, M. Mangano, P. Nason and L. Trentadue, CERN-TH/96-86,
   hep-ph/9604351.
\item \label{cdfpdf}
   J. Huston et al., MSU-HEP-50812, hep-ph/9511386;\\
   E.N. Glover, A.D. Martin, R.G. Roberts and W.J. Stirling, DTP/96/22,
   hep-ph/9603327.
\end{reflist}
 
\end{document}